\documentclass[useAMS,usenatbib]{mn2e}
\usepackage{natbib}
\usepackage{epsf}
\usepackage{graphicx}
\usepackage{textcomp}

\def\simlt{\mathrel{\rlap{\lower 3pt\hbox{$\sim$}}
        \raise 2.0pt\hbox{$<$}}}
\def\simgt{\mathrel{\rlap{\lower 3pt\hbox{$\sim$}}
        \raise 2.0pt\hbox{$>$}}}

\title[Jet breaks in black hole X-ray binaries]{Jet spectral breaks in black hole X-ray binaries}
\author[D.M. Russell et al.]{D.M. Russell$^{1,2,3}$\thanks{E-mail: russell@iac.es}, S. Markoff$^{3}$, P. Casella$^{4}$, A.G. Cantrell$^{5}$, R. Chatterjee$^{6}$,
\newauthor
R.P. Fender$^{7}$, E. Gallo$^{8}$, P. Gandhi$^{9}$, J. Homan$^{10}$, D. Maitra$^{8}$, J.C.A. Miller-Jones$^{11}$,
\newauthor
K. O'Brien$^{12}$, T. Shahbaz$^{1,2}$
\\
$^{1}$Instituto de Astrof\'isica de Canarias (IAC), V\'ia L\'actea s/n, La Laguna, E-38205, S/C de Tenerife, Spain\\
$^{2}$Departamento de Astrof\'isica, Universidad de La Laguna, La Laguna, E-38205, S/C de Tenerife, Spain\\
$^{3}$Astronomical Institute `Anton Pannekoek', University of Amsterdam, P.O. Box 94249, 1090 GE Amsterdam, the Netherlands\\
$^{4}$INAF - Osservatorio Astronomico di Roma, Via Frascati 33, I-00040 Monteporzio Catone (Roma), Italy\\
$^{5}$The Blake School, 511 Kenwood Pkwy, Minneapolis, MN 55403, USA\\
$^{6}$Department of Astronomy, Yale University, P.O. Box 208101, New Haven, CT 06520, USA\\
$^{7}$School of Physics and Astronomy, University of Southampton, Southampton, Hampshire SO17 1BJ, UK\\
$^{8}$Department of Astronomy, University of Michigan, 500 Church Street, Ann Arbor, MI 48109, USA\\
$^{9}$ISAS, Japan Aerospace Exploration Agency, 3-1-1 Yoshinodai, chuo-ku, Sagamihara, Kanagawa 229-8510, Japan\\
$^{10}$MIT Kavli Institute for Astrophysics and Space Research, 70 Vassar Street, Cambridge, MA 02139, USA\\
$^{11}$International Centre for Radio Astronomy Research - Curtin University, GPO Box U1987, Perth, WA 6845, Australia\\
$^{12}$Department of Physics, University of California, Santa Barbara, CA, USA\\
}
\begin{document}


\pagerange{\pageref{firstpage}--\pageref{lastpage}} \pubyear{2012}

\maketitle

\label{firstpage}

\begin{abstract}
In X-ray binaries, compact jets are known to commonly radiate at radio to infrared frequencies, whereas at optical to $\gamma$-ray energies, the contribution of the jet is debated. The total luminosity, and hence power of the jet is critically dependent on the position of the break in its spectrum, between optically thick (self-absorbed) and optically thin synchrotron emission. This break, or turnover, has been reported in just one black hole X-ray binary (BHXB) thus far, GX 339--4, and inferred via spectral fitting in two others, A0620--00 and Cyg X--1. Here, we collect a wealth of multiwavelength data from the outbursts of BHXBs during hard X-ray states, in order to search for jet breaks as yet unidentified in their spectral energy distributions. In particular, we report the direct detection of the jet break in the spectrum of V404 Cyg during its 1989 outburst, at $\nu_{\rm b} = (1.8 \pm 0.3) \times 10^{14}$ Hz ($1.7 \pm 0.2 \mu m$). 
We increase the number of BHXBs with measured jet breaks from three to eight. Jet breaks are found at frequencies spanning more than two orders of magnitude, from $\nu_{\rm b} = (4.5 \pm 0.8) \times 10^{12}$ Hz for XTE J1118+480 during its 2005 outburst, to $\nu_{\rm b} > 4.7 \times 10^{14}$ Hz for V4641 Sgr in outburst. A positive correlation between jet break frequency and luminosity is expected theoretically; $\nu_{\rm b} \propto L_{\rm \nu ,jet}^{\sim 0.5}$ if other parameters are constant. With constraints on the jet break in a total of 12 BHXBs including two quiescent systems, we find a large range of jet break frequencies at similar luminosities and no obvious global relation (but such a relation cannot be ruled out for individual sources). We speculate that different magnetic field strengths and/or different radii of the acceleration zone in the inner regions of the jet are likely to be responsible for the observed scatter between sources. There is evidence that the high energy cooling break in the jet spectrum shifts from UV energies at $L_{\rm X} \sim 10^{-8} L_{\rm Edd}$ (implying the jet may dominate the X-ray emission in quiescence) to X-ray energies at $\sim 10^{-3} L_{\rm Edd}$. Finally, we find that the jet break luminosity scales as $L_{\rm \nu ,jet} \propto L_{\rm X}^{0.56 \pm 0.05}$ (very similar to the radio--X-ray correlation), and radio-faint BHXBs have fainter jet breaks. In quiescence the jet break luminosity exceeds the X-ray luminosity.
\end{abstract}

\begin{keywords}
accretion, accretion discs, black hole physics, X-rays: binaries, ISM: jets and outflows
\end{keywords}

\section{Introduction}

The process of accretion onto compact objects is still not fully understood. How accretion leads to the production of relativistic, collimated jets, how much power is contained within these jets and how that power is distributed internally, are some of the major open questions in astronomy. Recently it has been established that stellar-mass black holes (BHs) can channel a large fraction of their accretion energy into these jets \citep*{gallet05,fend06,kordet06,tudoet06,russet07a}. These systems are X-ray binaries (XBs), in which the accreting matter is fed from a companion star via an accretion disc towards the BH.

A large scale height, poloidal magnetic field in the inner accretion flow is thought to launch the jet through magnetohydrodynamical processes, but it remains unclear how jets from accreting objects (Active Galactic Nuclei [AGN], XBs, $\gamma$-ray bursts) are formed and accelerated to relativistic velocities \citep*[likely by extraction of either the spin energy of the compact object, or the accretion energy from the accretion disc, or a combination; e.g.][]{blanzn77,blanpa82,liviet99,mckiet12}. For BHXBs, classical `flat spectrum' radio jets (like those seen in AGN) are commonly observed during hard X-ray states \citep[hereafter the hard state; for descriptions of X-ray states see][]{mcclet06,bell10}, when the accretion flow structure likely permits the existence of a large, vertical magnetic field. In softer X-ray states jets are observed to be quenched at radio frequencies \citep*[e.g.][]{corbet00,fend01,gallet03,russet11a} which may result from a suppression of the poloidal field by the geometrically thin disc which exists in the soft state \citep[e.g.][]{meie01}.

The flat, or slightly inverted ($\alpha \approx 0$--0.5 where $F_{\nu} \propto \nu^{\alpha}$) radio spectrum seen in the hard state extends to at least the millimetre regime \citep*{fendet00,fendet01,market01}, which can be explained by self-absorbed synchrotron emission from lepton populations at different radii from the BH; the signature of a conical, collimated jet \citep{blanko79,kais06}. At near-infrared (NIR) and higher frequencies, optically thin synchrotron emission has been detected \citep[e.g.][]{buxtba04,hyneet03,kaleet05,hyneet06,russet10,chatet11}, usually approximating a power law with spectral index $-1.0 \leq \alpha \leq -0.5$. This is expected from a compact jet spectrum at higher frequencies than the self-absorption \emph{break}, or \emph{turnover} in the spectral energy distribution \citep[SED;][]{blanko79,hjeljo88}. The jet origin of this optically thin synchrotron emission is supported by the extrapolation of the radio spectrum to NIR frequencies \citep*[e.g.][]{fend01,corbfe02,brocet04}, correlated radio and IR variability \citep[e.g.][]{fendet97,miraet98,eikeet98}, correlations with radio emission in the hard state \citep{russet06,coriet09}, quenching during the soft state \citep{jainet01,buxtba04,homaet05,buxtet12} and the recent detection of linear polarization \citep{shahet08,russfe08,russet11b}. The majority of the fast ($\sim$ seconds or less) variability reported at optical/NIR wavelengths is also likely to have a jet origin, since the variability has a spectrum consistent with optically thin synchrotron emission \citep[e.g.][]{hyneet03,hyneet06}, and is stronger at lower frequencies \citep[e.g.][]{gandet10,caseet10} which is inconsistent with both thermal emission and a nonthermal corona origin \citep*[for a recent nonthermal model see][]{veleet11}.

Physically, the size scale of the emitting region in the jet scales inversely with frequency. It is thought that the jet break frequency, $\nu_{\rm b}$ marks the start of the particle acceleration in the jet \citep*[e.g.][]{polket10}, at distances only $\sim 100 ~ r_{\rm g}$ (gravitational radii) from the BH \citep*[e.g.][]{market05,gallet07,miglet07,maitet09a,caseet10,peerma12}. Radiatively, the energetic output of the jet is dominated by the higher frequencies, and the peak flux density of the jet spectrum is at $\nu_{\rm b}$. In order to estimate the total power contained in the jet, to infer the fraction of accretion energy being channelled into these outflows, it is therefore necessary to identify the spectral break, and also the high energy synchrotron cooling break. The total radiative luminosity of the jet can only be inferred via $\nu_{\rm b}$, the luminosity at the break, $L_{\nu_{\rm b}}$ and the radiative efficiency, $\eta$. Measuring as accurately as possible the power contained in the jets at different luminosities is key to understanding the process of jet formation and the overall physics of accretion and the matter and energy XBs input into the interstellar medium (ISM). During outbursts, XBs can vary by eight orders of magnitude in luminosity, so it is possible to study how the accretion process and jet properties vary with mass accretion rate from luminosities $\sim 10^{-8}$--1 $L_{\rm Edd}$.

In addition to the total kinetic energy, other physical properties of the jet can be inferred by identifying the jet break. The cross section radius of the first acceleration zone (hereafter FAZ; i.e. the size of the region where the synchrotron power law starts) and magnetic field strength can be estimated directly \citep*[][]{rybili79,casepe09,chatet11}, while the velocity and opening angle can be inferred from fast timing fluctuations at the FAZ \citep{caseet10} and magnetic field ordering and orientation, via polarization at the FAZ \citep{shahet08,russfe08}. For jet models, observables like these are highly sought after. Constraining these parameters and how they vary with luminosity will dramatically improve attempts to model jets and simulate their production. Unlike in AGN, the rapid time dependency of XBs allows us to probe jet evolution, adding a further dimension to models \citep*[e.g.][]{maitet09b}.

Despite its importance, the jet break has only yet been observed directly in the spectrum of one BHXB (GX 339--4; \citealt{corbfe02,gandet11}; see also \citealt{nowaet05,coriet09}) and one neutron star XB \citep[4U 0614+09;][]{miglet06,miglet10} and inferred via spectral fitting in one BHXB \citep[Cyg X--1;][]{rahoet11}. In all three objects, the break is inferred to exist in the mid-IR, at $\sim (1-5)\times 10^{13}$ Hz, with GX 339--4 displaying large variability in $\nu_{\rm b}$ (by $> 1$ order of magnitude) on hour-timescales \citep{gandet11}. \cite{maitet11} show that the mid-IR spectrum of A0620--00 in quiescence is flat, and consistent with being self-absorbed synchrotron emission, while the NIR has a different power law, and is consistent with being optically thin synchrotron, which strongly favours a jet break between the two. The reasons why few jet breaks in XBs have been discovered to date are because the companion star or accretion disc can simply dominate the emission at these frequencies in some objects \citep[e.g.][]{miglet07,gallet07,rahoet11}, and few mid-IR data exist in the literature. Until recently, the only mid-IR detections of BHXBs in outburst at wavelengths $>$ 8$\mu$m were of GRO J0422+32, in which the source was detected at a level of 50 mJy at 11$\mu$m during a bright hard state \citep{vanpet94}.

Standard jet models predict the jet break frequency to depend on the mass accretion rate, black hole mass, location of the particle acceleration and magnetic field strength \citep*{falcbi95,heinsu03,market03,falcet04,chatet11}. If the latter three parameters are unchanged, a positive relation should exist between jet break frequency and mass accretion rate. Observational confirmation via any apparent correlation between break frequency and luminosity would provide strict constraints for jet models.

Here, we perform a comprehensive literature search for quasi-simultaneous multiwavelength (radio to optical) data of BHXBs in the hard state and quiescence \citep[we define quiescence as $L_{\rm X} < 10^{33.5}$ erg s$^{-1}$;][]{mcclet06}, in order to identify the jet break and test for a relation with luminosity. Specifically, we only gather data where there is evidence for synchrotron emission at NIR/optical frequencies. The data collection, SED construction and spectral fitting are described in Section 2. Each SED from every source is discussed, and we detail the method we use for isolating the jet emission in each case. The jet break is constrained in several sources, and the results are analysed in Section 3. The distribution of jet breaks and the global relation between jet break frequency and luminosity are analysed, and compared to relations expected theoretically. A large scatter in jet break frequency is found, and we discuss the possible origins of this scatter. A correlation between jet break luminosity and X-ray luminosity is also presented in Section 3, and we assess the likely contribution of the jet to the X-ray luminosity. A summary of the results are provided in Section 4.

\begin{table*}
\begin{center}
\caption{The data collected for this paper.}
\vspace{-2mm}
\begin{tabular}{lllllllll}
\hline
Source&Year&Dates&X-ray&$\Delta t$&$A_{\rm V}$&$D$  &$M_{\rm BH}$ &References\\
      &    &(MJD)&state&(d)       &(mag)      &(kpc)&($M_{\odot}$)&(data/parameters)\\
\hline
\emph{Black hole XBs:}               &            &           &         &       &                 &                 &                 &                 \\
GRO J0422+32                         & 1992       & 48874     & hard    & $< 1$ & $1.09 \pm 0.31$ & $2.49 \pm 0.30$ & $3.97 \pm 0.95$ & 1--4 / 5--6     \\
~~~~~(V518 Per)                      & 2000--10   & --        & quies.  & --    &                 &                 &                 & 6--8            \\
A0620--00                            & 1975       & 42648--50 & hard    & 2.0   & $1.05 \pm 0.12$ & $1.06 \pm 0.12$ & $6.61 \pm 0.25$ & 9--11 / 12      \\
~~~~~(V616 Mon)                      & 2005--6    & --        & quies.  & --    &                 &                 &                 & 12--14          \\
XTE J1118+480                        & 2000       & 51649     & hard    &$< 1^a$&$0.065\pm 0.020$ & $1.72 \pm 0.10$ & $8.53 \pm 0.60$ & 15--18 / 19     \\
~~~~~(KV UMa)                        & 2005       & 53386     & hard    & $< 1$ &                 &                 &                 & 20--22          \\
GS 1354--64                          & 1997       & 50772--4  & hard    & 3.0   & $2.60 \pm 0.31$ & $43 \pm 18$     & $> 7.33^b$      & 23--25 / 26--27 \\
~~~~~(BW Cir)                        &            &           &         &       &                 &                 &                 &                 \\
4U 1543--47                          & 2002       & 52490     & hard    & $< 1$ & $1.55 \pm 0.15$ & $7.5 \pm 0.5$   & $9.4 \pm 1.0$   & 28--29 / 30--31 \\
~~~~~(IL Lup)                        &            &           &         &       &                 &                 &                 &                 \\
XTE J1550--564                       & 2000       & 51697     & hard    & $< 1$ & $\sim 5.0$      & $4.38 \pm 0.58$ & $9.1 \pm 0.6$   & 32--34 / 35--38 \\
~~~~~(V381 Nor)                      & 2003       & 52750--1  & hard    & $2^f$ &                 &                 &                 & 39              \\
GX 339--4                            & 1997       & 50648     & hard    & $4^c$ & $3.25 \pm 0.50$ & 6 -- 15         & 4.3 -- 13.3     & 40--41 / 42--44 \\
~~~~~(V821 Ara)                      & 2010       & 55266     & hard    &$< 1^d$&                 &                 &                 & 42,45           \\
XTE J1752--223                       & 2010       & 55378     & hard    &$< 1^f$& $\sim 2.87$     & 3.5 -- 8        & $9.8 \pm 0.9$   & 46 / 46--48     \\
\multicolumn{2}{l}{~~~~~(SWIFT J1752.1-2220)}     &           &         &       &                 &                 &                 &                 \\
V4641 Sgr                            & 1999       & 51438     &soft?$^e$& $< 1$ & 0.775           & $5.5 \pm 2.5$   & $10.2 \pm 1.5$  & 49--50 / 49,51  \\
~~~~~(SAX J1819.3--2525)	     & 2002	  & 52419     & hard?	& $< 1$ &       	  &                 &                 & 53              \\
				     & 2003	  & 52857     & hard	& $< 1$ &       	  &                 &                 & 53--54          \\
MAXI J1836--194                      & 2011       & 55844--5  & hard    & $< 1$ & $1.31 \pm 0.23$ & $\sim 8^g$      & $\sim 10^g$     & 55 / 56         \\
Cyg X--1                             & 2005       & 53513     & hard    & $< 1$ & $\sim 2.95$     & $1.86 \pm 0.12$ & $14.8 \pm 1.0$  & 57 / 57--59     \\
~~~~~(V1357 Cyg)                     &            &           &         &       &                 &                 &                 &                 \\
V404 Cyg                             & 1989       & 47676     &soft?$^e$& $< 1$ & $4.0 \pm 0.4$   & $2.39 \pm 0.14$ & $8.8 \pm 0.4$   & 60--62 / 63--65 \\
\vspace{2mm}
~~~~~(GS 2023+338)                   & 1989       & 47728--9  & hard    & 1.1   &                 &                 &                 & 60,66--67       \\
\emph{Neutron star XB:}              &            &           &         &       &                 &                 &                 &                 \\
4U 0614+09                           & 2006       & 54038--42 & hard    & $< 4$ & 2.0             & $3.2 \pm 0.5$   & 1.4             & 68 / 68--70     \\
~~~~~(V1055 Ori)                     &            &           &         &       &                 &                 &                 &                 \\
\hline
\end{tabular}
\normalsize
\end{center}
The columns are: source name (alternative name), year, dates of observations, X-ray state, maximum time separation of the data, interstellar extinction, distance, BH mass and references (references of the data then for the parameters $A_{\rm V}$, $D$ and $M_{\rm BH}$). $^a$For the 2000 outburst of XTE J1118+480, the radio and sub-mm data are not strictly simultaneous, but \cite{fendet01} report a very steady source at these frequencies during the period MJD 51620--51720. $^b$Only a lower limit of $M_{\rm BH}$ is constrained for GS 1354--64. Here we assume a conservative $M_{\rm BH} < 30 M_{\odot}$ for the upper limit. $^c$For the 1997 SEDs of GX 339--4, one of the two radio fluxes was not quasi-simultaneous, but was calculated by \cite{corbfe02} from the well known radio--X-ray correlation. $^d$Some of the radio, NIR, optical and UV data in the SED of GX 339--4 in \cite{gandet11} straddled the date of the mid-IR data by six days either side, however the jet break we take here was measured from the mid-IR data only, which was strictly simultaneous. $^e$The radio spectrum is optically thin at this epoch, so the source was most likely not in a canonical hard state. $^f$No radio data were acquired for this epoch; as such only an upper or lower limit of the jet break frequency is inferred from the optical/IR data. $^g$The distance and BH mass of this BHXB are unknown; here we adopt typical values for a BHXB towards the Galactic centre, with conservative errors of a factor of four.
References:
(1) = \cite{vanpet94};
(2) = \cite{shraet94};
(3) = \cite*{kinget96};
(4) = \cite{goraet96};
(5) = \cite{hyne05};
(6) = \cite{geliha03};
(7) = \cite*{geliet10};
(8) = \cite{millet11};
(9) = \cite*{robeet76};
(10) = \cite*{kleiet76};
(11) = \cite{kuulet99};
(12) = \cite{cantet10};
(13) = \cite{maitet11};
(14) = \cite{gallet06};
(15) = \cite{chatet03a};
(16) = \cite{pavlet01};
(17) = \cite{tarash01};
(18) = \cite{fendet01};
(19) = \cite{geliet06};
(20) = \cite{hyneet06};
(21) = \cite{zuriet06};
(22) = \cite{brocet10};
(23) = \cite{castet97};
(24) = \cite*{soriet97};
(25) = \cite{brocet01};
(26) = \cite{casaet04};
(27) = \cite{casaet09};
(28) = \cite{buxtba04};
(29) = \cite{kaleet05};
(30) = \cite{oroset98};
(31) = \cite{oroset02};
(32) = \cite{jainet01};
(33) = \cite{corbet01};
(34) = \cite{russet10};
(35) = \cite*{tomset01};
(36) = \cite{tomset03};
(37) = \cite{kaaret03};
(38) = \cite{oroset11a};
(39) = \cite{chatet11};
(40) = \cite{corbfe02};
(41) = \cite{chatet02};
(42) = \cite{gandet11};
(43) = \cite{hyneet04};
(44) = \cite{shidet11};
(45) = \cite{cadoet11};
(46) = \cite{russet12};
(47) = \cite{rattet12};
(48) = \cite{shapet10};
(49) = \cite{chatet03b};
(50) = \cite{hjelet00};
(51) = \cite{oroset01};
(52) = \cite{uemuet04a};
(53) = \cite{uemuet04b};
(54) = \cite*{rupeet03};
(55) = \cite{russet11c};
(56) = Russell et al. (2013; in preparation);
(57) = \cite{rahoet11};
(58) = \cite{reidet11};
(59) = \cite{oroset11b};
(60) = \cite{hanhj92};
(61) = \cite*{johnet89};
(62) = \cite{casaet91};
(63) = \cite{hyneet09};
(64) = \cite{millet09};
(65) = \cite*{kharet10};
(66) = \cite*{gehret89};
(67) = \cite{leibet91};
(68) = \cite{miglet10};
(69) = \cite{kuulet10};
(70) = \cite{vanset00}.
\end{table*}

\section{Data collection and analysis}

In addition to the published jet breaks identified in the SEDs of GX 339--4, Cyg X--1 and 4U 0614+09 \citep{corbfe02,miglet10,gandet11,rahoet11}, we conducted a literature search for radio to optical SEDs of BHXBs. In order to identify the jet break the following criteria were imposed: (1) The BHXB must be in a hard state or in quiescence at the time; (2) There must be evidence for synchrotron emission at frequencies $\nu > 10^{12}$ Hz (i.e., several orders of magnitude higher in frequency than radio), with a measurable spectral index; (3) There must be at least one radio data point; (4) All hard state data must be quasi-simultaneous -- taken within one day, or a few days at most (those with time separations $> 2.0$ d are discussed individually below).

In some works, constraints (upper or lower limits, but not direct measurements) of the jet break frequency have been made by measuring the optical/NIR spectral index to either be optically thick or optically thin, but without quasi-simultaneous radio data \citep[e.g.][]{chatet11}. We include these upper/lower limits in our analysis, but in order to measure the jet break frequency itself (not an upper/lower limit) we require at least one quasi-simultaneous radio data point. For sources in quiescence, quasi-simultaneity is not required since fluxes are thought not to vary considerably (although as we will see, the jet break frequency can shift on short timescales; we discuss this caveat applicable to these quiescent data in our analysis). Nevertheless, for A0620--00 different optical `quiescent states' exist \citep[][see Section 2.2]{cantet10}. The broad H$\alpha$ emission line can be very prominent in the optical spectrum \citep[e.g.][]{casaet91,fendet09}; this line resides within the $R$-band filter (centred at $6400$\AA). In some SEDs the $R$-band data point appeared high compared to the fluxes in other filters at similar wavelengths. These $R$-band data points were removed. Quasi-simultaneous X-ray fluxes were also acquired.

Evidence for synchrotron emission in all cases is spectral. The classical example is GX 339--4, where a well documented `V'-shape SED is evident; the red component originating in synchrotron emission from the jet (which is quenched in the soft state), and the blue component which is possibly from the irradiated accretion disc \citep{corbfe02,homaet05,coriet09,cadoet11,shidet11,buxtet12,rahoet12,dincet12}. As a visual example, the reader is directed to the SEDs of GX 339--4 presented in \cite{corbfe02} and \cite{gandet11}. Brightness temperature arguments, variability which is stronger at lower frequencies (including cross-correlations with X-ray) and polarization confirm the jet origin of the red component \citep{caseet10,gandet10,gandet11,russet11b}. The red component normally dominates the NIR flux during the hard state \citep[e.g.][]{russet06}, but in some BHXBs it appears fainter or absent, compared to the blue component \citep[e.g.][]{hyneet02,soleet10}. For this study it is important to find data where the synchrotron emission is not only present, but it can be isolated from the blue component and its spectral index can be measured. This is necessary in order to extrapolate its power law towards lower frequencies. Where there is evidence for disc emission also in the optical/NIR SED, this contribution has been subtracted, leaving just the synchrotron spectrum. Methods of subtraction are explained in the individual subsections for each source. Where two spectral components are evident with clear, different spectral indices, we take just the reddest bands to measure the spectral index (incorporating the uncertainty into the errors). Other methods to subtract the disc flux and isolate the jet emission include measuring the disc flux in each band from its exponential decay in the soft state \citep{russet10,russet12,dincet12} and by measuring its rapid variability in several bands simultaneously \citep{hyneet03,hyneet06}.

Ten BHXBs were found with SEDs satisfying the above criteria, in addition to the two BHXBs mentioned above with jet breaks already reported. The 12 sources, dates and references are given in Table 1, as are the best known values for their interstellar extinctions, distances and BH masses. The neutron star XB 4U 0614+09, with a published jet break, is also tabulated. All optical/NIR data were de-reddened adopting the extinction curve of \cite*{cardet89}. Mid-IR data were de-reddened using the relation of \cite{weindr01}.

After isolating the flux of the synchrotron emission, we construct the SEDs. Unless the jet break itself is clearly visible in the SED, we fit power laws to the radio and IR/optical SEDs in order to infer the jet break frequency and its flux. If the synchrotron power law at $\nu > 10^{12}$ Hz has a spectral index $\alpha \leq -0.4$, this is consistent with optically thin synchrotron and the best fit power law and its errors are extrapolated to lower frequencies. Likewise, the best fit radio power law is extrapolated to higher frequencies and the jet break is defined by interpolating these two power laws. We take the error on the jet break to be the most extreme outcomes using the upper/lower limits to each of the two power law fits. The resulting frequency ranges are shown by horizontal double-ended arrows in each panel of Fig. 1.

This method effectively assumes the jet spectrum can be approximated by a broken power law, which is the classical picture \citep{blanko79}. More complex SEDs may be more appropriate in some cases, such as an additional excess of emission in the SED at around the jet break frequency \citep{peerca09,market05} but most of our SEDs can be well fit by a broken power law (see Section 3 for more discussion on this). In this sense, by interpolating we are measuring the position of the `classical' jet break in most cases, which may lie under this excess. In addition, re-brightenings due to internal shocks in the jet, or other processes, may introduce excess emission above the `flat' optically thick spectrum \citep[for example there was a millimetre excess in the optically thick spectrum of XTE J1118+480;][]{market01}. Using interpolation between radio and IR/optical power laws is therefore an approximation of the jet break between the optically thick and thin emission, but may not represent the true peak flux of the jet, if the SED is more complex than this. There are few sub-mm data available in the literature, as most sub-mm telescopes are barely sensitive to detect these mJy sources. However, this is now changing, with upgrades to current sub-mm telescopes and new instrumentation that are significantly more sensitive than before (e.g. the James Clerk Maxwell Telescope, the Submillimeter Array, the Stratospheric Observatory for Infrared Astronomy and the Atacama Large Millimeter Array). Given the available data until now, we find that in most cases the SEDs can be well approximated by a broken power law (see Fig. 1), but in a follow-up work we will model the SEDs using state of the art jet models. For the purposes of this paper, we simply wish to constrain the jet break (as defined by the interpolation of radio and IR/optical synchrotron power laws) in many sources and test for a general relation with luminosity.

X-ray fluxes are converted to bolometric luminosities using the approximation $L_{\rm bol} \approx 5 \times L_{\rm X,2-10 keV}$ (unless bolometric luminosities were quoted in the papers) for hard state objects \citep[which has an associated error of $\leq 10$ per cent;][]{miglfe06} and assuming an X-ray power law of index $\Gamma = 1 - \alpha = 1.6$ in the hard state and $\Gamma = 2.0$ in quiescence \citep*[fairly typical values; e.g.][]{corbet06}. Since the same power law index is assumed for every source (except in quiescence), the bolometric correction is the same for each source, so the original 2--10 keV fluxes are proportional to the bolometric luminosities. When X-ray fluxes are absorbed, we use the NASA tool $\emph{WebPIMMS}$\footnote{Available at http://heasarc.nasa.gov/Tools/w3pimms.html} and the known values of hydrogen column density to obtain unabsorbed fluxes. The bolometric luminosity in Eddington units is calculated from the distance and BH mass estimates given in Table 1. The errors in the distance and BH mass are propagated into the error in the bolometric luminosity (we take the full ranges of each to infer the total possible range of values of luminosities). The SEDs and the jet breaks derived for each source are discussed in the following subsections.

\begin{figure*}
\centering
\includegraphics[width=5.8cm,angle=270]{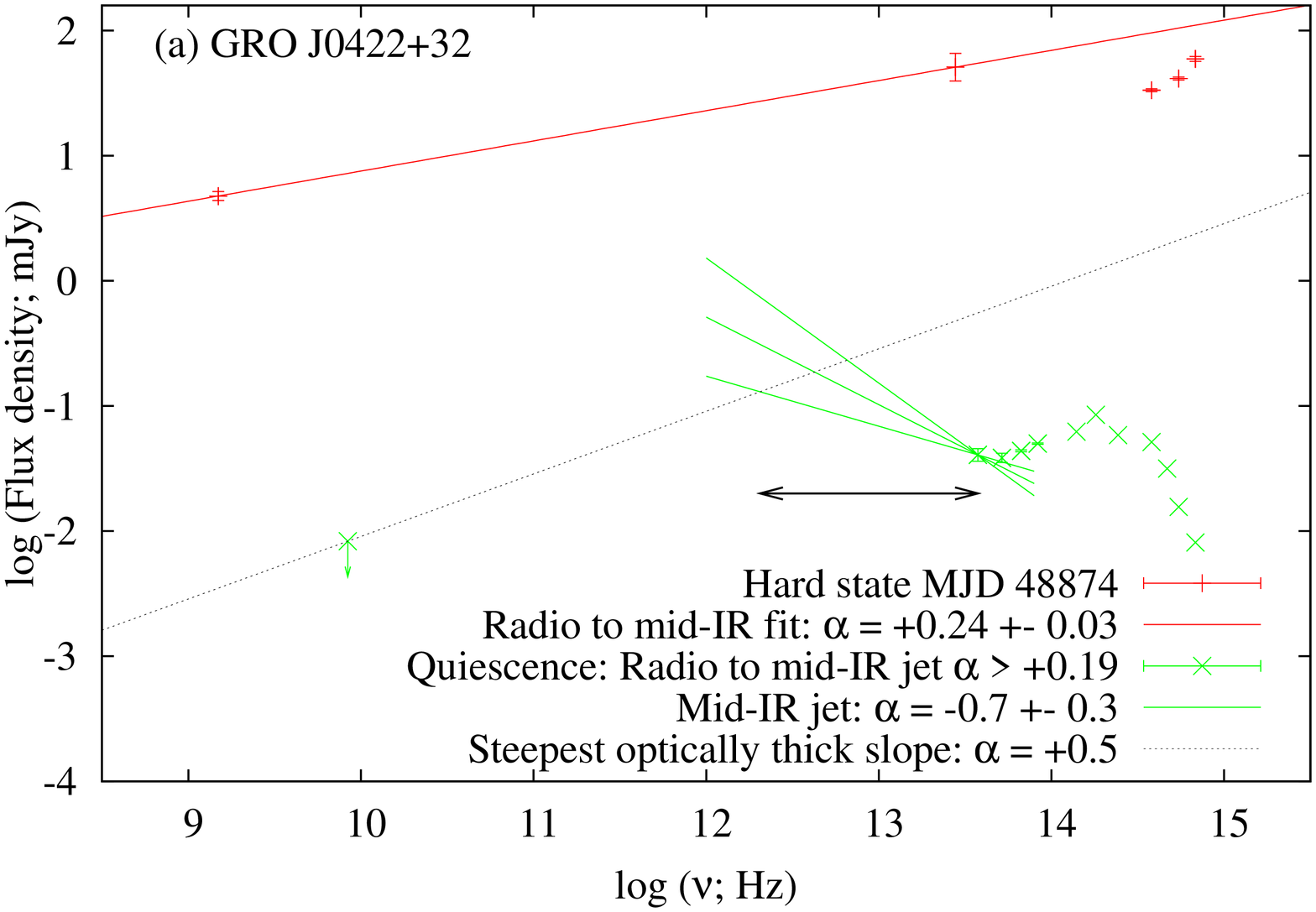}
\includegraphics[width=5.8cm,angle=270]{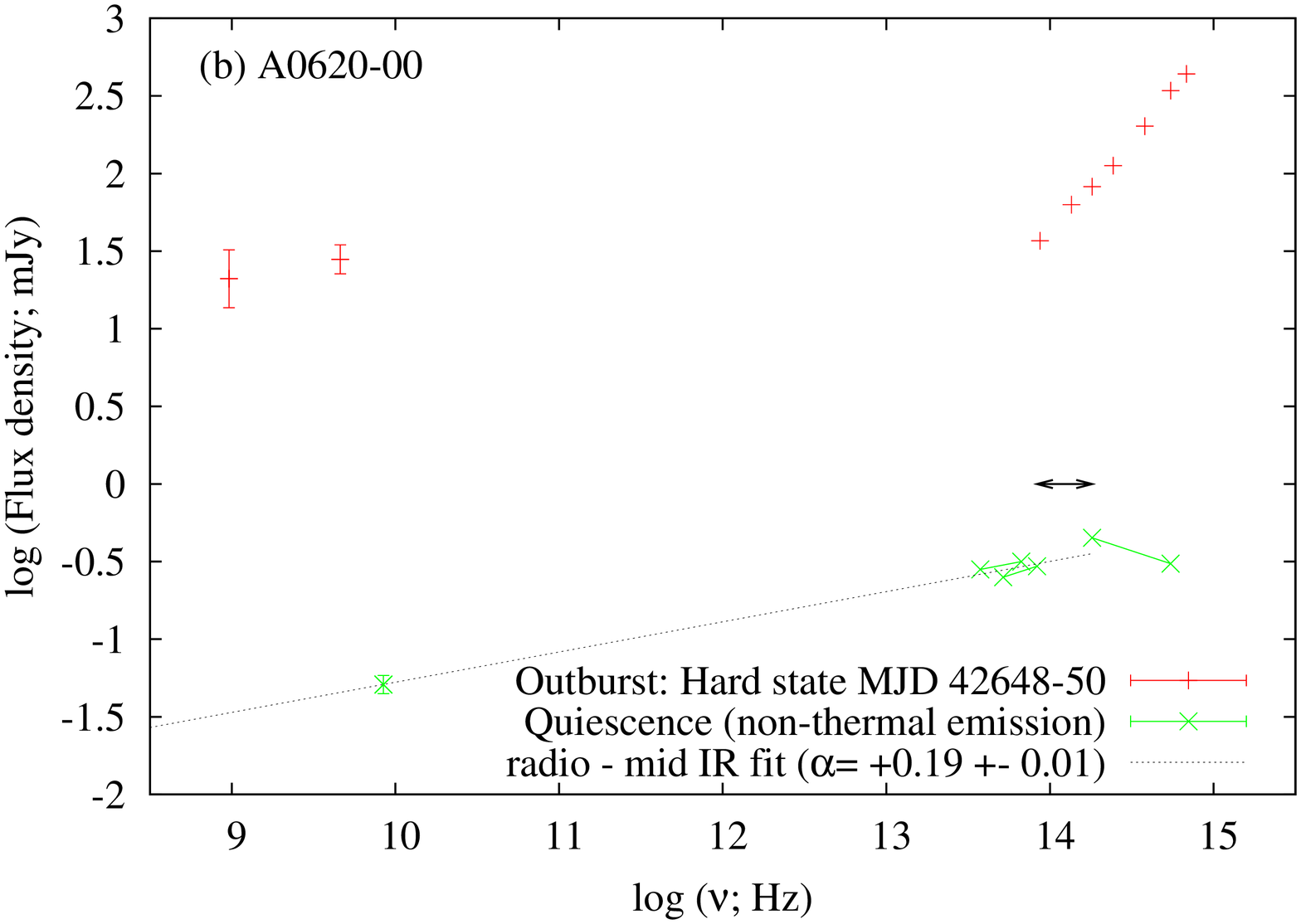}\\
\includegraphics[width=5.8cm,angle=270]{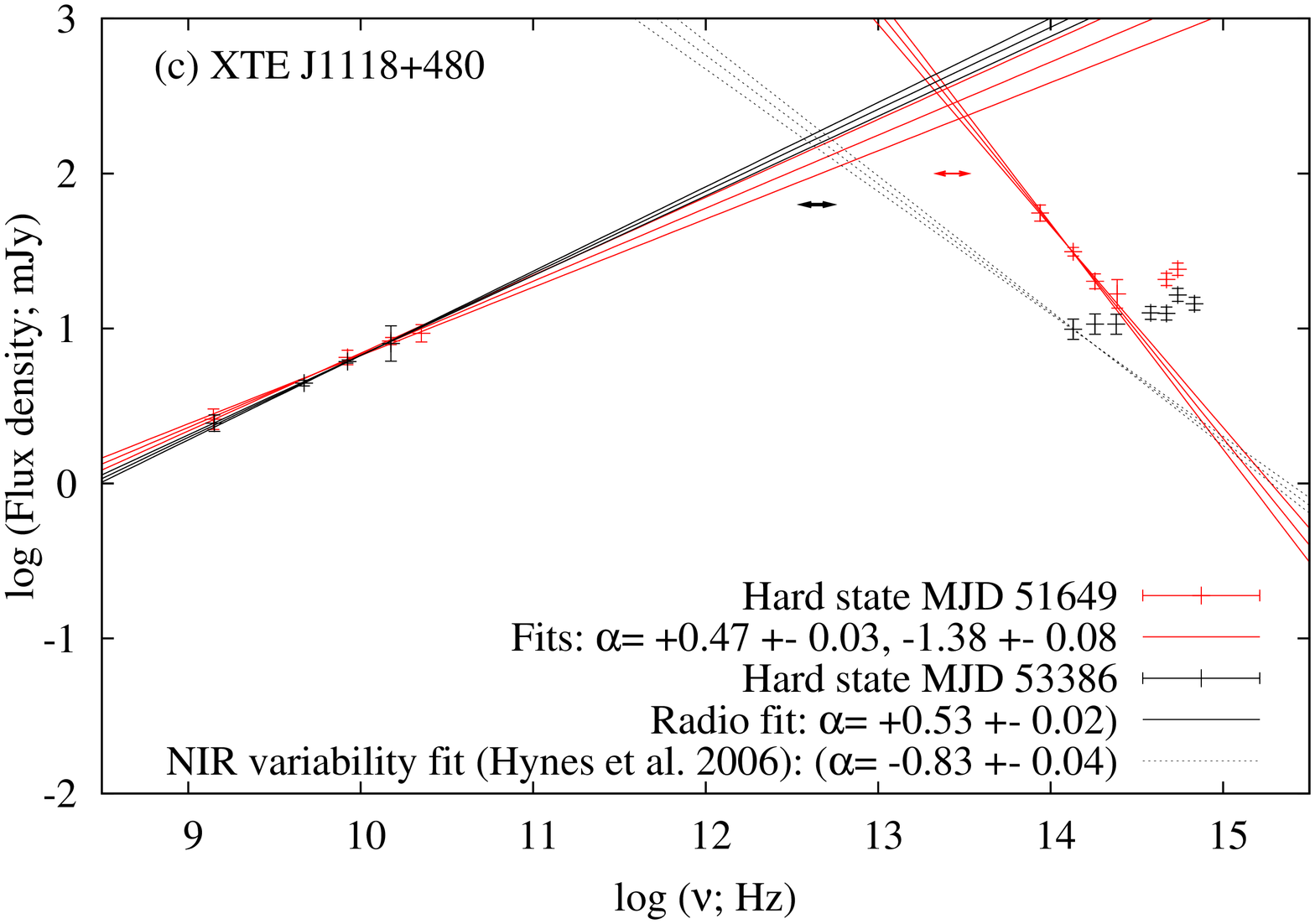}
\includegraphics[width=5.8cm,angle=270]{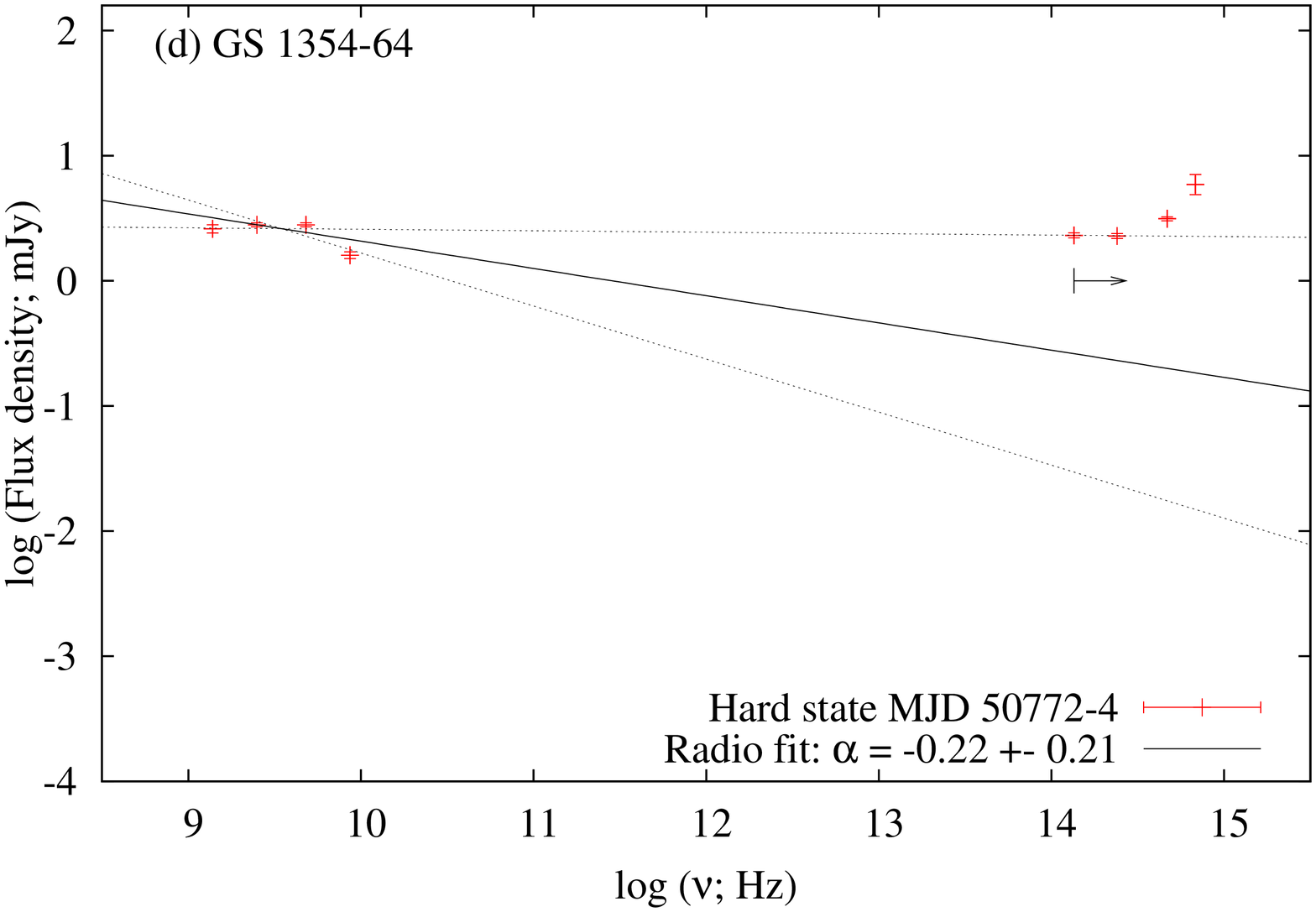}\\
\includegraphics[width=5.8cm,angle=270]{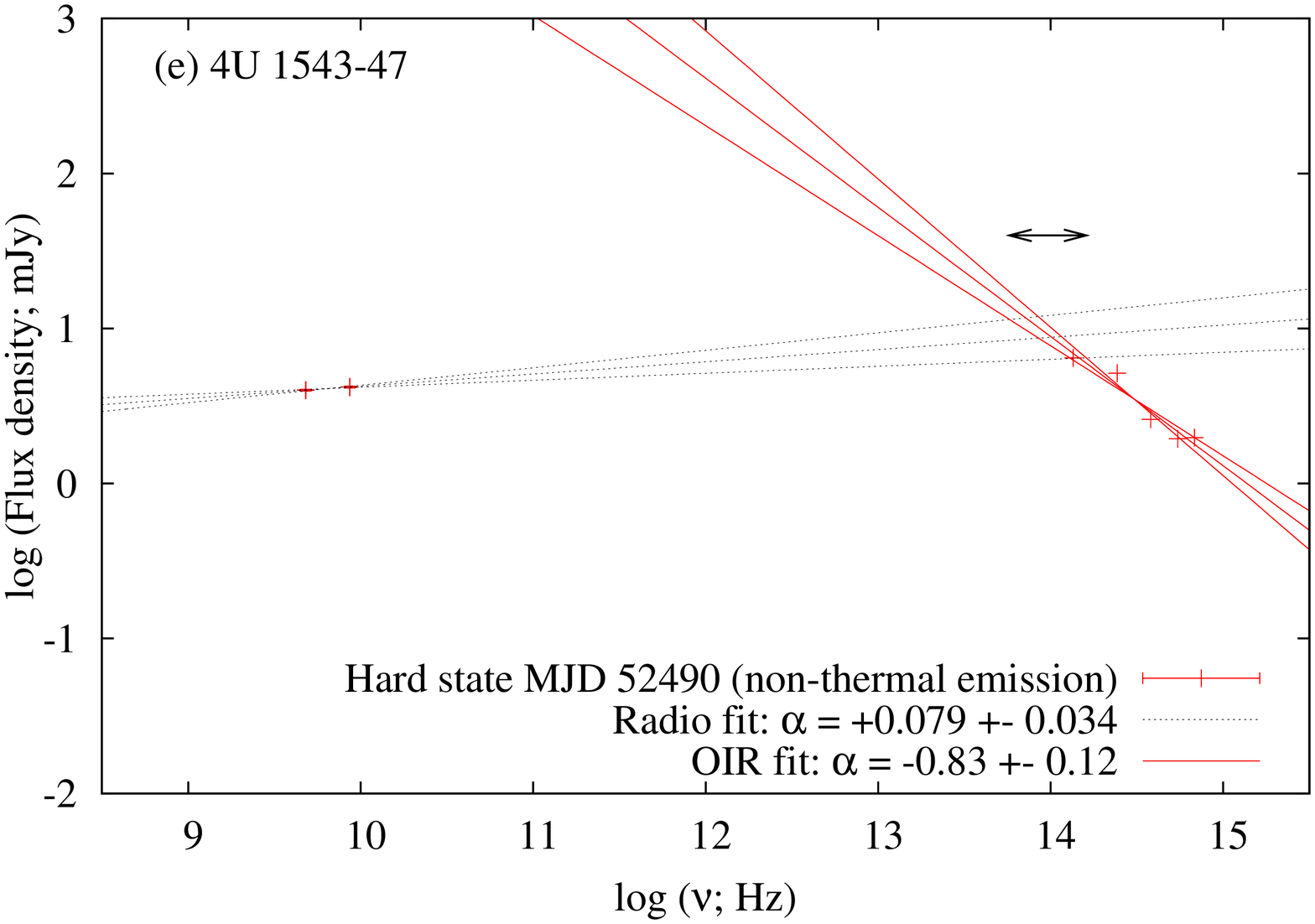}
\includegraphics[width=5.8cm,angle=270]{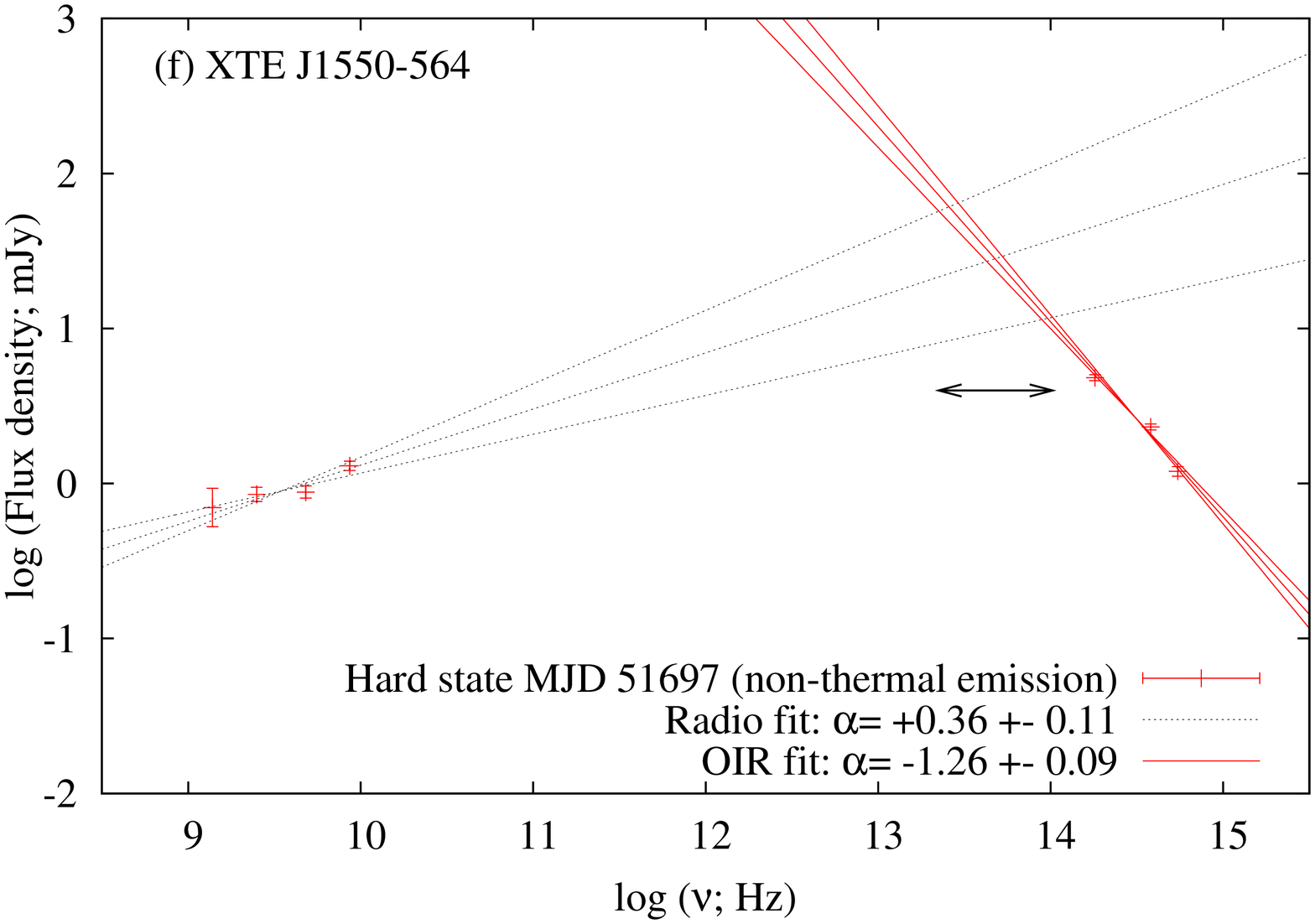}\\
\includegraphics[width=5.8cm,angle=270]{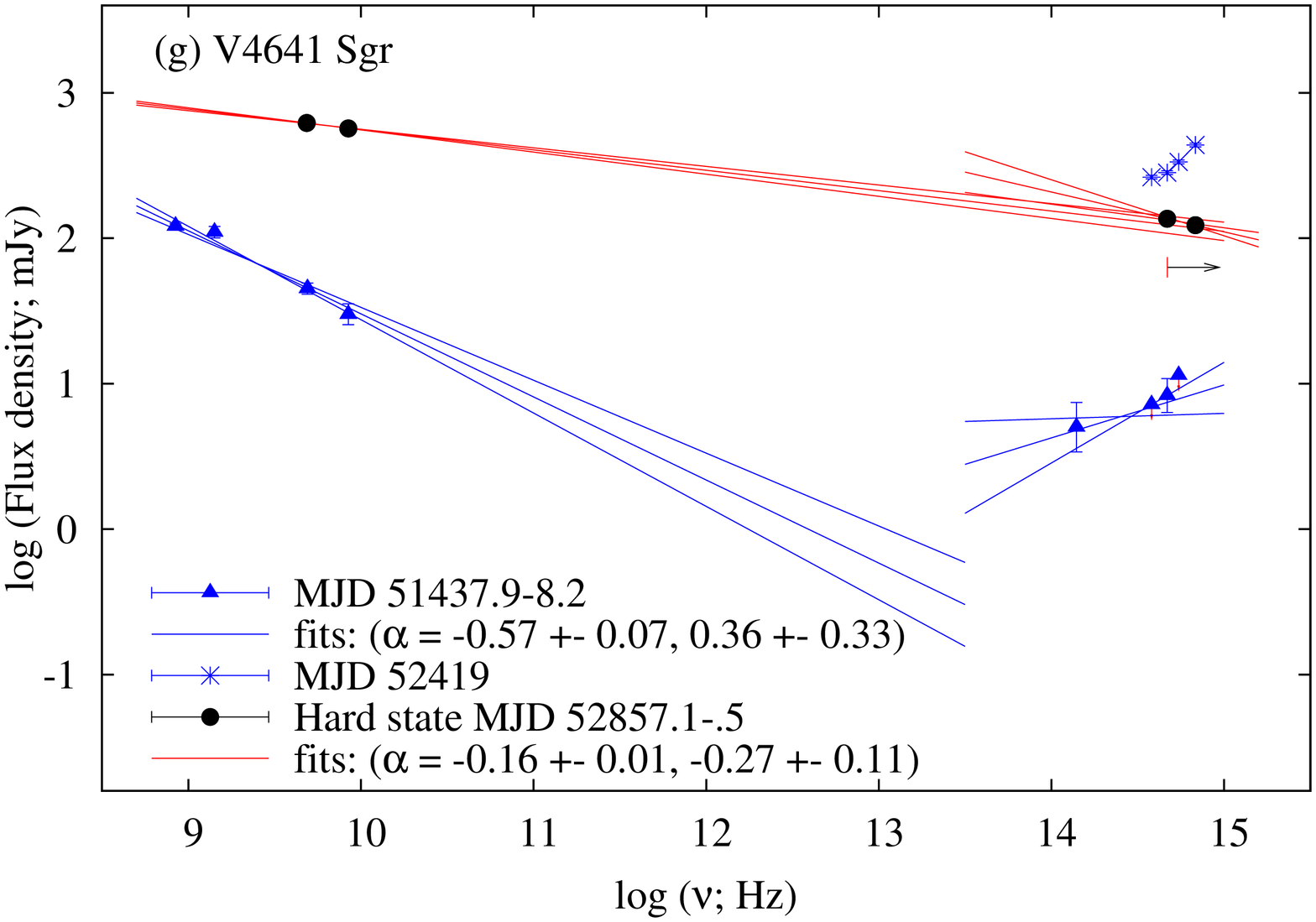}
\includegraphics[width=5.8cm,angle=270]{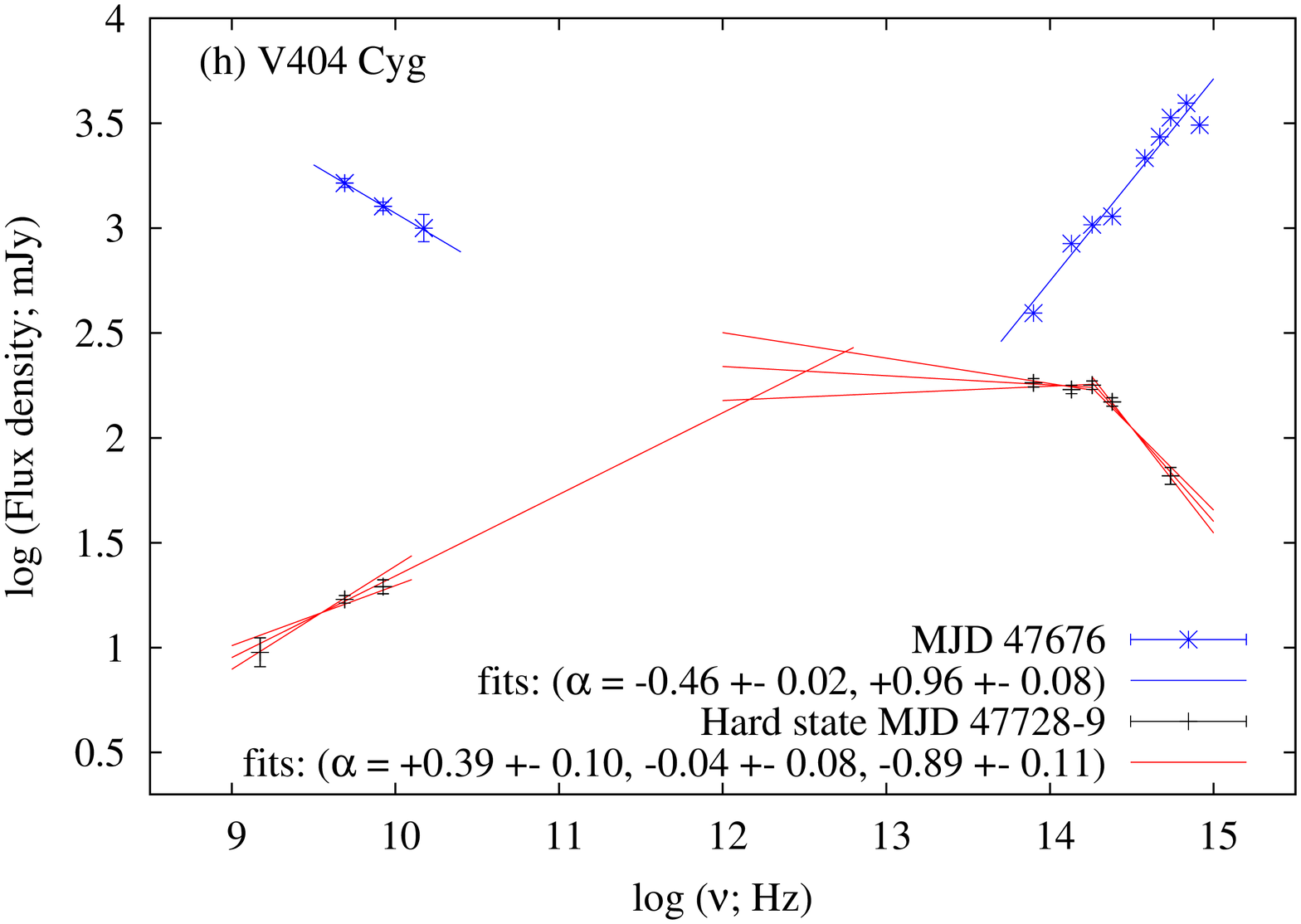}\\
\caption{Radio to optical SEDs of eight BHXBs, with power law fits to various regions of the spectra used to constrain the jet break flux and frequency (see text for details). Data from the hard (red and black), quiescent (green) and unknown (blue) states are shown.}
\end{figure*}

\subsection{GRO J0422+32}

A wealth of optical photometry was acquired during the 1992 hard state outburst of this transient BH, but the spectrum was fairly blue, with no reported evidence for synchrotron emission. No NIR data were taken, but \cite{vanpet94} reported a bright, $51 \pm 9$ mJy detection at 10.8$\mu m$ in the mid-IR during outburst which later faded to $< 36$ mJy (3$\sigma$ upper limit) at 10.2$\mu m$. The accretion disc, secondary star and heated dust were all ruled out as the origin \citep{vanpet94}. An X-ray driven accretion disc wind was the only plausible explanation, but this would require more mass to be lost via the wind than accreted onto the BH. Such strong winds have been detected but not in the hard state, when winds are generally found to be suppressed \citep{neille09,pontet12}.

In hindsight it seems viable, and expected, that this mid-IR detection could be synchrotron emission from the jet. \cite{shraet94} monitored the source at radio frequencies during the outburst and found a decaying radio source with an inverted spectrum; $\alpha > 0$. A radio observation was made within 1 d of the bright mid-IR detection; its flux was $4.8 \pm 0.4$ mJy at 1.49 GHz. Optical magnitudes were also reported on the same day by \cite{kinget96} and \cite{goraet96}. In Fig. 1a the (de-reddened) radio, mid-IR and optical detections from this epoch are shown by red crosses. The radio to mid-IR spectral index is $\alpha = +0.24 \pm 0.03$, which is fairly typical of self-absorbed synchrotron from the jet, and similar to the spectral index of the radio spectrum seen during this outburst \citep{shraet94}. The reddest optical band ($I$-band, centred at $7900$\AA) on the same date was $33.4 \pm 0.9$ mJy, so the spectrum must be $\alpha < 0$ between mid-IR and optical, implying that a turnover is necessary in the jet spectrum at lower frequencies than $I$-band ($\nu_{\rm b} < 3.8 \times 10^{14}$ Hz). We therefore propose the jet as the most likely source of the bright mid-IR detection. Since the optically thin spectral index cannot be measured, we can place no further constraint on the jet break frequency or the flux at the peak. The jet break itself could lie at frequencies above or below the mid-IR detection.

Optical, NIR, mid-IR and radio data of GRO J0422+32 have been acquired during quiescence. \cite{geliet10} showed that the optical to mid-IR SED is best fit by the companion star plus an optically thin synchrotron jet with spectral index $\alpha \approx -0.7$ (the jet produces $\sim$ all the 8 $\mu m$ emission). This spectral index is inconsistent with a dusty circumbinary disc origin, which has a bluer ($\alpha > 0$) SED at these wavelengths \citep{munoma06}. A deep radio observation achieved a stringent 3$\sigma$ upper limit of $8.3 \mu$Jy at 8.4 GHz in quiescence \citep{millet11}. The SED is shown as green crosses in Fig. 1a. The radio upper limit and mid-IR jet detection give a radio to mid-IR jet spectral index of $\alpha > +0.19$. This is not unexpected, since during outburst some radio spectra were quite inverted, with $\alpha > +0.2$ \citep{shraet94}. No radio spectrum from a BHXB has been reported steeper than $\alpha \sim +0.5$, so we consider this as an upper limit. The black dotted line in Fig. 1a illustrates this maximum optically thick jet flux extrapolated to higher frequencies. We adopt a conservative error on the optically thin spectral index of $\alpha = -0.7 \pm 0.3$. This extrapolated to lower frequencies is shown as green lines in Fig. 1a. From where the extrapolated mid-IR jet spectrum errors meet the maximum possible optically thick jet flux, we show that the jet break frequency must be located at $\nu_{\rm b} > 2.1 \times 10^{12}$ Hz. The jet break cannot be at higher frequencies than 8 $\mu m$ ($\nu_{\rm b} < 3.7 \times 10^{13}$ Hz) because the emission is optically thin here. The black arrow in Fig. 1a indicates the range of possible jet break frequencies constrained here. In this case we can also constrain the peak flux at the jet break. The maximum flux is set by the upper limit of $\alpha = -1.0$ for the optically thin spectrum crossing the maximum flux from the optically thick spectrum. We take the observed 8 $\mu m$ flux itself as the minimum flux at the jet break. \cite{millet11} also tabulate the quiescent X-ray luminosity of GRO J0422+32 to be $3.2 \times 10^{30}$ erg s$^{-1}$ (3--9 keV), which equates to a bolometric luminosity (see above) of $L_{\rm X} = 2.3 \times 10^{31}$ erg s$^{-1}$, or $L_{\rm X} = (7.1 \pm 2.7) \times 10^{-8} L_{\rm Edd}$.

\subsection{A0620--00}

For a brief time at the start of the 1975 outburst of A0620--00, this BHXB was in a hard X-ray state, with flat spectrum radio emission \citep[][and references therein]{kuulet99}. Optical to NIR data were taken; the source was detected in a total of 7 photometric bands (up to $L$-band at 3.5 $\mu m$) within two days of the radio data (Table 1). In Fig. 1b the red crosses represent this epoch. It is clear that no jet emission is evident in the optical/NIR data; the SED is consistent with a power law; $\alpha = +1.2$, typical of an irradiated disc \citep[e.g.][]{hyne05}. The radio to NIR $L$-band spectral index is flat, but no constraint can be made on the jet break frequency.

In quiescence, A0620--00 has been detected at radio frequencies by \cite{gallet06}. A mid-IR excess is known to exist above the companion star and accretion disc emission in quiescence, and it has been speculated that this could be due to the jet or a circumbinary disc \citep{munoma06,gallet07}. \cite{maitet11} present mid-IR and NIR/optical data of the source during quiescence. During quiescence A0620--00 exhibits `passive' and `active' states \citep{cantet10}, and \cite{maitet11} note that the mid-IR data were taken during an `active' state. After subtracting the stellar light from the companion star and the accretion disc flux as measured in the `passive' state, the remaining flux is nonthermal and variable \citep{maitet11}. The 3.6--8.0 $\mu m$ nonthermal spectrum is consistent with a power law of index $\alpha = +0.2$ in one observation and $\alpha = +0.3$ in another. This is inconsistent with the spectrum of a dusty circumbinary disc \citep{munoma06}, although we cannot rule out a circumbinary disc making a weak contribution. During the `active' state the optical--NIR nonthermal spectrum has a spectral index; $\alpha = -0.7 \pm 0.2$ \citep{cantet10,maitet11}. This is consistent with optically thin synchrotron.

In Fig. 1b we plot the quiescent SED (green crosses) from the radio and mid-IR data, and add $V$-band (centred at $5500$\AA) and $H$-band ($1.66 \mu m$) data from the epoch in the active state close in time to the mid-IR data \citep{maitet11}. It was already mentioned in \cite{maitet11} that the break between optically thin and optically thick synchrotron seems to lie between the optical/NIR and the mid-IR regimes, from the observed spectral indices. Here, we show that the radio to mid-IR spectral index is $\alpha = +0.19 \pm 0.01$, and the radio detection is consistent with the extrapolation of the measured mid-IR synchrotron spectral index. This, assuming no additional components so far unconsidered are present, further supports the claim that the mid-IR nonthermal emission is very likely to originate in the jet. The jet break is measured here to lie between 3.6 $\mu m$ and 1.7 $\mu m$ ($\nu_{\rm b} = (1.3 \pm 0.5) \times 10^{14}$ Hz). This is the first time the jet break has been inferred for a BHXB in quiescence.

\subsection{XTE J1118+480}

Two SEDs were acquired of this BHXB, one from each of its two hard state outbursts. The 2000 outburst in particular had excellent multiwavelength coverage (this halo BHXB lies behind a very low level of extinction, and UV spectroscopy was possible). While the radio spectrum remained steady for 100 days \citep{fendet01}, NIR photometry revealed a bright, red SED (as bright as $L = 8.7$ mag, or 0.1 Jy at 3.5 $\mu m$). On 2000 April 15, data in four NIR bands and two optical bands were acquired, and result in a `V'-shape SED (Fig. 1c; red crosses). The optical bands appear to be blue ($\alpha > 0$) while the NIR bands are well fit by a power law with spectral index $\alpha = -1.38 \pm 0.08$. Although this is fairly steep for optically thin synchrotron emission, the spectral index of the jet of XTE J1550--564 was also seen to evolve over several days from a steeper one than seen here, to a value typical of optically thin synchrotron emission \citep{russet10}. The steeper index may be indicative of a thermal, possibly Maxwellian distribution of electrons at this time \citep{russet10}. The steady radio spectrum is also well fit by a power law of index $\alpha = +0.47 \pm 0.03$. By propagating the errors in these two power law fits we constrain the range of frequencies and fluxes where they must meet. We derive a jet break frequency from this of $\nu_{\rm b} = (2.8 \pm 0.6) \times 10^{13}$ Hz. At a different epoch during this outburst, the NIR flux appeared much flatter, and a constraint on the jet break could not be made directly, but was inferred via spectral modelling to be around $\sim 4 \times 10^{14}$ Hz \citep{market01,maitet09a}.

Radio, NIR and optical data were acquired on the same day during the 2005 hard state outburst of XTE J1118+480 (Table 1). This time the optical/NIR SED was slightly blue ($\alpha > 0$) but \cite{hyneet06} obtained \emph{strictly simultaneous} NIR $J$, $H$, $K$ fast photometry. By measuring the flux of the variable component in each filter, \cite{hyneet06} found that the rapidly variable component had a red spectrum consistent with optically thin synchrotron (on 2005 January 16 it was $\alpha = -0.83 \pm 0.04$). The observed optical/NIR and radio fluxes (at four radio frequencies) are plotted in Fig. 1c (black crosses), while the spectral fit to the variable NIR component is shown (black dotted lines), as is the power law fit to the radio data. Similarly to the SED during the 2000 outburst, we are able to interpolate the NIR and radio synchrotron power laws to infer $\nu_{\rm b}$. Here in the 2005 SED, the radio spectrum has a very similar flux and spectral index to the 2000 SED, whereas the NIR flux and spectral index are quite different, resulting in a jet break at $\nu_{\rm b} = (4.5 \pm 0.8) \times 10^{12}$ Hz, a frequency almost one order of magnitude lower than for the 2000 SED. This demonstrates that even though the radio spectra are very similar, the total radiative power of the jet can vary substantially due to the changing conditions near the jet base.

\begin{figure}
\centering
\includegraphics[width=6cm,angle=270]{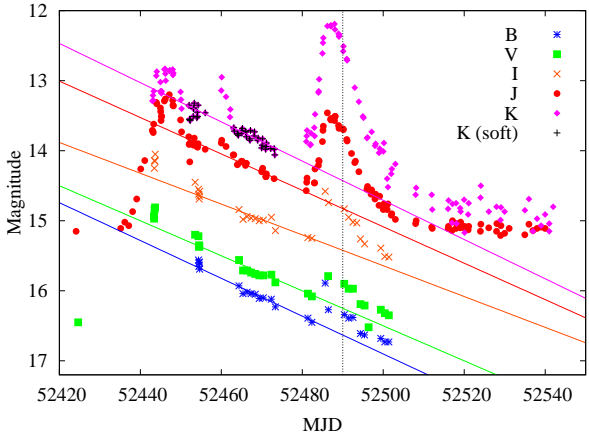}
\caption{Optical/NIR light curve of the decay of the 2002 outburst of 4U 1543--47 \citep{buxtba04}. The re-brightening in the decaying hard state is clearly visible after MJD 52480. Exponential decays are fitted to the soft state light curves. As an example the $K$-band (centred at $2.2 \mu m$) data fitted is shown as black crosses \citep[a brief flare during the soft state is not included in the fit;][]{buxtba04}. The excess flux above the exponential decay is measured at the time of the radio observation (marked by the vertical dotted line).}
\end{figure}

\subsection{GS 1354--64}

In 1997 this BHXB exhibited a hard state outburst. Evidence for a NIR excess above the disc spectrum was reported in \cite{brocet01}, and they point out that its origin is likely to be synchrotron because the flat radio spectrum extrapolates to the same level of flux as this excess in the NIR. Here, we take data observed on MJD 50772--4, when radio, NIR and optical data were acquired. The NIR ($J$- to $K$-band; 1.2--2.2 $\mu m$) SED is flat; too flat ($\alpha \approx 0$) for optically thin synchrotron emission even after subtracting any disc contamination. The radio SED has a power law of index $\alpha = -0.22 \pm 0.21$ on this date which, extrapolated to the NIR is consistent with the flat NIR SED (see Fig. 1d; note that the highest frequency radio data point is lower than the others, which may imply a more complex SED. If this point is removed the radio SED is much flatter). The jet break must therefore lie at frequencies higher than the $K$-band (2.2 $\mu m$).

\subsection{4U 1543--47}

Synchrotron emission in the NIR/optical has been seen to quench during the soft state and return in the hard state in this BHXB \citep{buxtba04,kaleet05,russet07b}. During the soft state of the 2002 outburst, the light curve could be described by an exponential decay \citep{buxtba04}. Radio observations were performed during the hard state decline of the outburst on MJD 52490. The radio spectrum appeared flat/slightly inverted \citep{kaleet05}. Here, we use the published optical/NIR light curves to subtract the disc component from the optical/NIR flux on the date of the radio observations. Fig. 2 shows the light curve, with exponential decay fits to the soft state data in each band (see figure caption for details). This method of extrapolating the disc flux measured in the soft state into the hard state was successfully adopted and used to measure the jet spectral index in the decays of XTE J1550--564 \citep{russet10}, XTE J1752--223 \citep{russet12} and GX 339--4 \citep{dincet12}. The vertical dotted line indicates the date of the radio observations; MJD 52490. From these fits, we measure a fractional disc contribution to the observed flux of 76\%, 72\%, 58\%, 35\% and 17\% in the $B$-, $V$-, $I$-, $J$ and $K$-bands (0.44--2.2 $\mu m$), respectively on the same date as the radio observation.

In Fig. 1e the de-reddened radio to optical SED of the synchrotron emission is shown, after the disc emission has been subtracted. The optical/NIR SED has a spectral index of $\alpha = -0.83 \pm 0.12$, typical of optically thin synchrotron. The interpolation of the flat radio and optically thin optical/NIR power laws infers possible jet break frequencies at $\nu_{\rm b} = (1.1 \pm 0.5) \times 10^{14}$ Hz.

\subsection{XTE J1550--564}

This source was monitored regularly at optical/NIR wavelengths during its 2000 outburst \citep{jainet01}. A strong component from synchrotron emission was evident at these wavelengths during the hard state, which disappeared in the soft state and reappeared in the hard state decline \citep{jainet01,russet07b,russet10}. Although few radio observations were made during this outburst, \cite{corbet01} report a 4-band observation on one date shortly after transition to the fading hard state. The radio spectrum was inverted, consistent with a compact jet. The thermal contribution to the optical/NIR data in the fading hard state was isolated from the jet synchrotron emission by \cite{russet10}, by subtracting the exponential decay of the thermal flux in each filter. In Fig. 1f the SED of the jet on the date of the radio observation is shown. This is similar to fig. 8 in \cite{russet10} but here, the thermal flux has been subtracted. The spectral index of the optical/NIR jet flux on this date was $\alpha = -1.26 \pm 0.09$. The jet spectral index evolved from this moderately steep value (possibly representing a thermal distribution of electrons) to values of $\alpha \sim -0.7$ (more typically of optically thin synchrotron) during the initial phase of the hard state decay \citep[see fig. 2 of][]{russet10}. By interpolating the power law fits to radio and optical/NIR jet emission, we find that the jet break must lie at a frequency of $\nu_{\rm b} = (6.3 \pm 4.0) \times 10^{13}$ Hz at this time.

An optical/NIR SED of XTE J1550--564 from its 2003 hard state outburst was presented in \cite{chatet11}. The de-reddened SED (see their fig. 3) is consistent with optically thin synchrotron, but a slight flattening at the lowest NIR frequencies implies the jet break may be around $\sim 10^{14}$ Hz at this time, but could exist at lower frequencies. It cannot reside at higher frequencies than $J$-band ($1.3 \mu m$). No radio data were taken at this epoch so here we adopt an upper limit of the jet break frequency, of $\nu_{\rm b} \leq 2.4 \times 10^{14}$ Hz ($J$-band). Since the $Ks$-band ($2.2 \mu m$) flux is likely to be close to the peak flux, we adopt a peak jet flux of 1--2 times the observed de-reddened $Ks$-band flux.

\subsection{GX 339--4}

The first claim of a detection of a jet break in a BHXB was in \cite{corbfe02}, where the SED of GX 339--4 showed a characteristic reduction in flux in the $J$-band (centred at $1.3 \mu m$) compared to $H$- and $K$-bands (1.7 and $2.2 \mu m$) in the NIR. The NIR flux level was consistent with the extrapolation of the radio power law, during a hard state. We include this SED in our analysis, and refer the reader to figs. 1 and 2 of \cite{corbfe02} for the SED.

More recently, \cite{gandet11} found a dramatically variable jet break from time-resolved mid-IR data taken with the \emph{Wide-field Infrared Survey Explorer (WISE)} satellite during a bright hard state in 2010. Thirteen mid-IR SEDs (4 bands $simultaneously$ observed within the 3.4--22 $\mu m$ wavelength range) of the source within 24 hours indicated high-amplitude hour-timescale variability, including shifts in the jet break frequency by one order of magnitude at least through this wavelength range. Here we take the jet break frequency and flux values from epochs 12 and 13, where the jet break was seen directly in the mid-IR SED \citep[fig. 3;][]{gandet11}. In our analysis we also indicate the full range of possible jet break frequencies measured from the $WISE$ data. The broadband SED, including radio, NIR, optical, UV and X-ray data is presented in fig. 1 of \cite{gandet11}.

\subsection{XTE J1752--223}

There is evidence for synchrotron emission contributing to the optical/NIR flux of XTE J1752--223 during its 2010 outburst \citep{curret11,russet12}. During the outburst decay, the light from the disc and jet were separated using the extrapolation of the exponential decay of the disc flux as measured in the soft state \citep{russet12}. The optical spectral index of the jet emission varied during the decay. On one date near the end of the outburst (MJD 55378), optical and NIR data were acquired and a spectral index of the jet could be measured; $\alpha = -1.0 \pm 0.3$, which is consistent with optically thin synchrotron emission \citep[this SED of the jet is shown in green in fig. 7 of][]{russet12}. No radio data were taken on the same date, so we cannot measure the jet break frequency directly. However, the jet break must exist at a frequency lower than $H$-band ($1.7 \mu m$) for the optical/NIR SED to be optically thin (the $Ks$-band $2.2 \mu m$ disc-subtracted flux is not well constrained). The peak jet flux density cannot be measured accurately. We take its lower limit as the $H$-band ($1.7 \mu m$) flux lower limit, and its upper limit as two times the $H$-band upper limit. The radio flux at this time is likely to be less than 0.3 mJy because the radio source was decaying \citep[see e.g.][]{rattet12} and this was its flux two weeks prior to this epoch. If the peak jet flux in IR was brighter than two times the $H$-band upper limit, this would produce a very inverted radio-to-IR spectrum (the radio-to-optical spectral index was measured to be $\alpha \sim +0.05$ two weeks before this). We therefore constrain the jet break to be at $\nu_{\rm b} \leq 1.8 \times 10^{14}$ Hz with a peak flux density of 0.36--2.28 mJy.

\subsection{V4641 Sgr}

This BHXB had a very bright, rapid outburst in 1999, followed by a number of fainter ones several years later. Optical flares seen from this BHXB have been proposed to originate from synchrotron emission \citep{uemuet04a,uemuet04b}. Radio, NIR and optical data were taken on one date (MJD 51437.9--51438.2) during the 1999 outburst, and we present this SED as blue solid triangles in Fig. 1g (see Table 1 for data references). The companion star in this system is fairly bright, and we subtract the known phase-dependent companion star flux from the total flux in each band \citep{chatet03b}. Here, the radio spectrum is optically thin, and the source was not in the canonical hard state at the time. The optical--NIR (non-stellar) SED is blue and is not close to the extrapolated radio jet power law.

Optical and radio data were also acquired on the same date, MJD 52857 during the 2003 outburst of this source. Although the optical flux varied rapidly, \cite{uemuet04b} presented simultaneous $B$ and $R$-band data ($4400$\AA and $6400$\AA). The source is bright in both optical and radio at this time, and the companion star only contributes $\sim 10$ per cent of the optical flux. The non-stellar optical SED has a slightly negative spectral index this time, and is not consistent with thermal emission. The radio spectral index within 0.4 days of the optical observation was measured very accurately \citep{rupeet03}, and extrapolates to exactly the level of the optical flux (black circles in Fig. 1g). We note that on a different date, MJD 52419 in 2002, a four-band optical SED was obtained and shows a much bluer, slightly brighter non-stellar spectral slope (blue crosses in Fig. 1g; no radio data were available on this date and the X-ray spectral state is uncertain). If thermal emission dominated the optical flux on MJD 52857, we would expect its SED to be blue like it is on MJD 52419. Instead, the optical spectral index ($\alpha = -0.27 \pm 0.11$ is consistent with being the same as the radio spectral index ($\alpha = -0.16 \pm 0.01$) within errors, and the optical flux lies on the accurately measured extrapolated radio power law. We therefore interpret this optical emission as originating in the jet. The SED favours the scenario of the jet break residing at higher frequencies than the optical bands, but the optical spectral index is $< 3 \sigma$ away from values expected from optically thin synchrotron, so we cannot rule out this possibility. However, the jet break cannot be at frequencies lower than the optical $R$-band because the optical jet emission would not lie on the extrapolated radio power law if this were the case. We can therefore constrain the jet break frequency to be higher than the frequency of $R$-band ($6400$\AA) on MJD 52857; $\nu_{\rm b} \geq 4.7 \times 10^{14}$ Hz. On the other two dates, we cannot constrain the jet break from these data.

\subsection{MAXI J1836--194}

This BHXB, discovered in 2011, was detected brightly in the NIR--mid-IR regime (2--12 $\mu m$) using VISIR on the VLT, when it was in the hard state \citep{russet11c}. On MJD 55845 the source was brightest, with a 12 $\mu m$ detection of $57 \pm 1$ mJy and the spectral index was measured to be consistent with optically thin synchrotron emission. The jet break must therefore reside at a lower frequency than this mid-IR band; $\nu_{\rm b} \leq 2.5 \times 10^{13}$ Hz.

\subsection{Cyg X--1}

Cyg X--1 is a high-mass X-ray binary and a persistent source that usually resides in the hard state but occasionally performs state transitions to a softer state. \cite{rahoet11} present mid-IR Spitzer spectra on three dates in different states. Although the SED is dominated by the bright O star companion, the flux is found to be variable between dates. They fit the spectrum using a model of the stellar continuum, the compact jet and an additional power law from bremsstrahlung emission from the wind of the O star. For their observation 1, the best fit (with the lowest reduced $\chi ^2$) is with a broken power law from the compact jet (fixing the optically thin spectral index to $\alpha = -0.6$), with a jet break frequency of $\nu_{\rm b} = $ (2.70 -- 2.94) $\times 10^{13}$ Hz. The jet break is not directly visible in their spectrum, and the raw jet spectrum (after subtracting the stellar and bremsstrahlung contributions) is not presented. Instead the jet break is inferred via spectral fitting by assuming a broken power law with zero curvature. We include this claimed jet break in our analysis. It is relevant to note that in GX 339--4 (and in AGN) the break is smoothly curved over a factor of at least a few in frequency \citep{gandet11}.

\subsection{V404 Cyg}

During its 1989 outburst, V404 Cyg was a bright X-ray, optical and radio source. A 3-band radio SED and a 9-band optical--NIR SED (spanning $> 1$ order of magnitude in frequency) were acquired almost simultaneously on MJD 47676 near the outburst peak (see Table 1 for data references). The radio and de-reddened optical--NIR flux densities exceeded 1 Jy, and the brightest optical/NIR reddened magnitudes were $V = 12.2$; $K = 7.7$; $L' = 7.2$ (0.05, 0.55 and 0.33 mJy, respectively). The radio SED is typical of optically thin synchrotron at this time, and indeed the optical--NIR SED is blue and can be described by a power law of index $\alpha = 1$, with no evidence for any IR excess (blue crosses in Fig. 1h). The source was not likely to have been in the canonical hard state at this time.

Later in the outburst the radio spectrum evolved to a flat/inverted one typical of a hard state compact jet \citep{hanhj92} and the X-ray spectrum was hard. On MJD 47728--47729 a 5-band optical--NIR SED was obtained within 1.1 days of a 3-band radio SED (see Table 1 for data references). On this date the radio spectrum was inverted, and the optical--NIR SED was red, and inconsistent with a single power law (black crosses in Fig. 1h). The optical--IR SED was well fitted by a broken power law; a flat ($\alpha = -0.04 \pm 0.08$) SED in the IR joining a redder ($\alpha = -0.89 \pm 0.11$) SED in the optical regime. The SED can be well described by a jet break. The self-absorbed synchrotron regime is breaking to optically thin synchrotron around the $H$-band ($1.7 \mu m$) in this SED. The optical--NIR SED cannot be fitted by a blackbody and cannot be explained by thermal emission. We infer a jet break frequency of $\nu_{\rm b} = (1.8 \pm 0.3) \times 10^{14}$ Hz.

Interestingly, the SED requires a slight curvature, or a second break between radio and IR. This curvature has been seen in some compact jets of AGN and is also consistent with theoretical SEDs of compact jets produced by some models and simulations \citep*[e.g.][see also above]{peerca09,jamiet10}. The existence of curvature or a broken power law describing the optically thick spectrum has implications for our method of assuming a single power law for most sources studied here. The possible effects this has on our results is discussed in Section 3.2.3.

V404 Cyg also has a well sampled quiescent SED, with radio, mid-IR, NIR, optical, UV and X-ray fluxes measured \citep{munoma06,gallet07,hyneet09}. The companion star dominates the whole mid-IR to UV SED but an excess at 24 $\mu m$ has a similar flux density to the quiescent flat radio spectrum. The spectral index of this excess could not be measured, and its origin is unclear, so we cannot constrain the jet break in V404 Cyg in quiescence. Note that the flux of the companion star produces a negligible amount of flux in the above SEDs from the 1989 outburst.

\subsection{The neutron star source 4U 0614+09}

After the first claim of the direct detection of a jet break in a XB, in GX 339--4 \citep{corbfe02}, 4U 0614+09 was the next, and the first (and to date the only) secure detection of the jet break in the SED of a neutron star XB \citep{miglet06,miglet10}. The source is persistent, and normally resides in the hard state. \cite{miglet10} collected quasi-simultaneous radio, mid-IR, NIR, optical, UV and X-ray data and discovered the jet break exists between two Spitzer bands (8 and 24 $\mu m$), at $\nu_{\rm b} = $(1.25 -- 3.71)$ \times 10^{13}$ Hz. We add this jet break to our analysis in order to compare our BH sample with a neutron star XB.

\begin{figure*}
\centering
\includegraphics[width=8.3cm,angle=0]{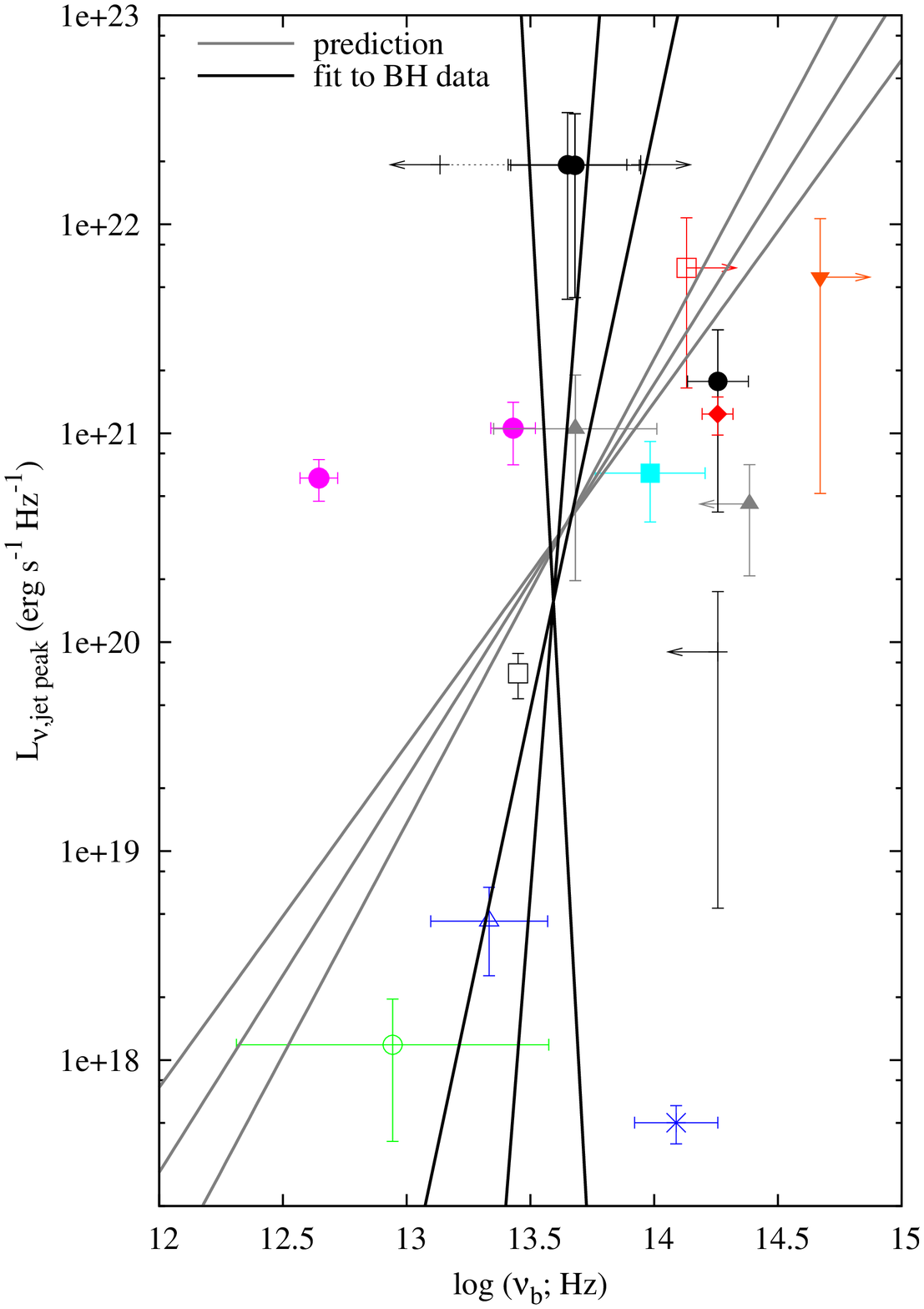}
\includegraphics[width=8.3cm,angle=0]{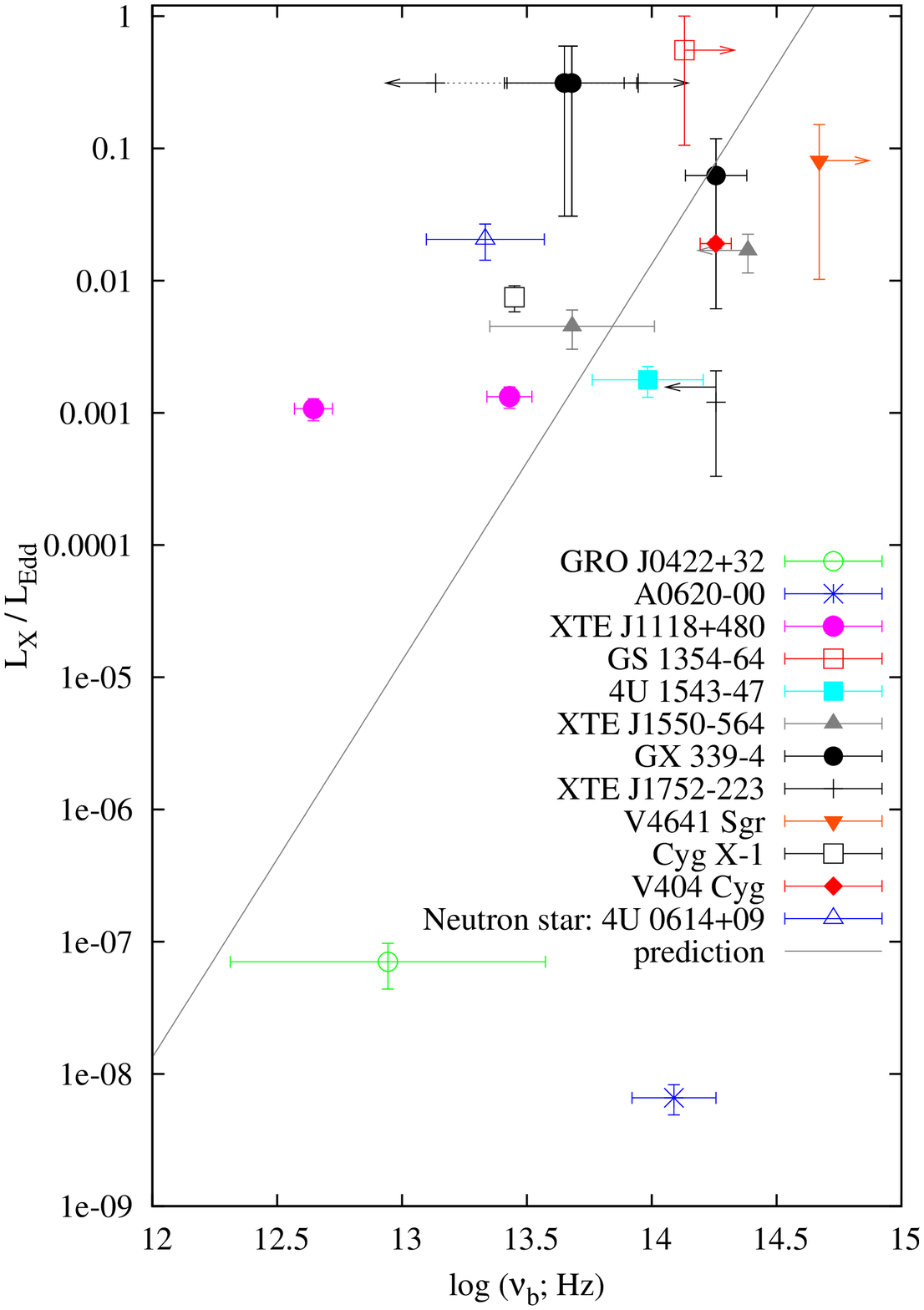}
\caption{Jet break frequency versus jet break monochromatic luminosity (left panel) and X-ray luminosity as a fraction of the Eddington luminosity (right). The predicted relations (and their errors) as defined by equations (10) and (4) are shown, as is the best power law fit to the black hole data in the left panel (equation (13)).}
\end{figure*}

\begin{table*}
\begin{center}
\caption{Table of results: constraints on the jet spectral break.}
\vspace{-6mm}
\begin{tabular}{llllllll}
\hline
Source&MJD&$\alpha_{\rm thick}$&$\alpha_{\rm thin}$&log ($\nu_{\rm b}$; Hz)&$L_{\rm \nu ,jet}$     &$L_{\rm X} / L_{Edd}$&Ref.\\
      &   &			&		     &  		     &(erg s$^{-1}$Hz$^{-1}$)&                   &	   \\
\hline
\emph{Black hole XBs:}          &           &         &       &                 &                 &                 &                 \\
GRO J0422+32    & --	    & $\leq +0.5$     &$-0.70 \pm 0.30$&$12.94 \pm 0.63$&$(1.18 \pm 0.78) \times 10^{18}$&$(7.06 \pm 2.67) \times 10^{-8}$& 1 \\
A0620--00       & --	    & $+0.19 \pm 0.01$&$-0.70 \pm 0.20$&$14.09 \pm 0.17$&$(5.01 \pm 1.04) \times 10^{17}$&$(6.60 \pm 1.70) \times 10^{-9}$& 2 \\
XTE J1118+480   & 51649     & $+0.47 \pm 0.03$&$-1.38 \pm 0.08$&$13.43 \pm 0.09$&$(1.06 \pm 0.35) \times 10^{21}$&$(1.33 \pm 0.25) \times 10^{-3}$& 1 \\
XTE J1118+480   & 53386     & $+0.53 \pm 0.02$&$-0.83 \pm 0.04$&$12.65 \pm 0.08$&$(6.11 \pm 1.37) \times 10^{20}$&$(1.07 \pm 0.20) \times 10^{-3}$& 1 \\
GS 1354--64     & 50772--4  & $-0.22 \pm 0.21$& --	       &$> 14.13$	&$(6.19 \pm 4.55) \times 10^{21}$&$(5.53 \pm 4.47) \times 10^{-1}$& 1 \\
4U 1543--47     & 52490     & $+0.08 \pm 0.03$&$-0.83 \pm 0.12$&$13.98 \pm 0.22$&$(6.45 \pm 2.69) \times 10^{20}$&$(1.78 \pm 0.47) \times 10^{-3}$& 1 \\
XTE J1550--564  & 51697     & $+0.36 \pm 0.11$&$-1.26 \pm 0.09$&$13.68 \pm 0.33$&$(1.05 \pm 0.85) \times 10^{21}$&$(4.50 \pm 1.48) \times 10^{-3}$& 1 \\
XTE J1550--564  & 52750--1  & --	      & --	       &$< 14.39$	&$(4.58 \pm 2.50) \times 10^{20}$&$(1.69 \pm 0.55) \times 10^{-2}$& 3 \\
GX 339--4       & 50648     & $+0.08 \pm 0.08$& --	       &$14.26 \pm 0.12$&$(1.77 \pm 1.35) \times 10^{21}$&$(6.23 \pm 5.62) \times 10^{-2}$& 4 \\
GX 339--4       & 55266$^a$ & $+0.29 \pm 0.02$&$-0.73 \pm 0.24$&$13.65 \pm 0.24$&$(1.93 \pm 1.49) \times 10^{22}$&$(3.12 \pm 2.81) \times 10^{-1}$& 5 \\
GX 339--4       & 55266$^b$ & ''	      & ''	       &$13.68 \pm 0.26$&$(1.91 \pm 1.47) \times 10^{22}$& ''				  & 5 \\
GX 339--4       & 55266$^c$ & ''	      & ''	       &$< 13.13$	&$>4.53 \times 10^{21}$ 	 & ''				  & 5 \\
GX 339--4       & 55266$^d$ & ''	      & ''	       &$> 13.95$	&$>2.06 \times 10^{21}$ 	 & ''				  & 5 \\
XTE J1752--223  & 55378     & --	      &$-1.00 \pm 0.30$&$< 14.26$	&$(9.00 \pm 8.46) \times 10^{19}$&$(1.20 \pm 0.87) \times 10^{-3}$& 6 \\
V4641 Sgr       & 52857     & $-0.16 \pm 0.01$& --	       &$> 14.67$	&$(5.58 \pm 5.07) \times 10^{21}$&$(8.07 \pm 7.05) \times 10^{-2}$& 1 \\
MAXI J1836--194 & 55844--5  & --	      &$-0.68 \pm 0.02$&$< 13.40$	& --$^e$			 & --$^e$			  & 7 \\
Cyg X--1        & 53513     & $+0.01 \pm 0.02$& --	       &$13.45 \pm 0.02$&$(7.10 \pm 1.73) \times 10^{19}$&$(7.47 \pm 1.65) \times 10^{-3}$& 8 \\
\vspace{2mm}
V404 Cyg        & 47728--9  &$+0.39\pm 0.10^f$&$-0.89 \pm 0.11$&$14.26 \pm 0.06$&$(1.24 \pm 0.26) \times 10^{21}$&$(1.91 \pm 0.15) \times 10^{-2}$& 1 \\
\emph{Neutron star XB:}     &           &         &       &                 &                 &                 &                 \\
4U 0614+09      & 54038--42 & $+0.03 \pm 0.04$&$-0.47 \pm 0.15$&$13.33 \pm 0.24$&$(4.62 \pm 2.09) \times 10^{18}$&$(2.05 \pm 0.63) \times 10^{-2}$& 9 \\
\hline
\end{tabular}
\normalsize
\end{center}
The columns are: source name, dates of observations, optically thick synchrotron spectral index, optically thin synchrotron spectral index, log(jet break frequency), monochromatic luminosity at the jet break, bolometric luminosity as a fraction of the Eddington luminosity, and references. $^{a-d}$These refer to WISE observations 12, 13, 4 and 8 respectively, as given in \cite{gandet11}. $^e$The distance to MAXI J1836--194 is unconstrained at this time, so luminosities cannot yet be calculated. $^f$The optically thick spectrum of V404 Cyg was best fit by a broken power law; $\alpha_{\rm thick} = +0.39 \pm 0.10$ at radio frequencies and $\alpha_{\rm thick} = -0.04 \pm 0.08$ at infrared.
References:
(1) This paper;
(2) \cite{maitet11};
(3) \cite{chatet11};
(4) \cite{corbfe02};
(5) \cite{gandet11};
(6) \cite{russet12};
(7) \cite{russet11c};
(8) \cite{rahoet11};
(9) \cite{miglet10}.
\end{table*}

\section{Results and analysis}

We have measured directly, constrained, or inferred upper/lower limits on the jet break frequency in nine BHXBs. Together with constraints from three additional BHXBs reported in the literature and one neutron star source, we can for the first time study the distribution of jet break frequencies and test for a relation between jet break frequency and luminosity. From our spectral fits we infer the range of possible peak jet fluxes (i.e. the flux at the jet break) and we take published quasi-simultaneous X-ray luminosities (for data references see Table 1). In Table 2 we present our results; all jet break frequencies, luminosities, optically thick and thin power law indices, and X-ray luminosities are tabulated.

In the following subsections we compare the distribution of jet breaks and the global relation between jet break frequency and luminosity in the context of the relations expected theoretically. We also test for a global correlation between the luminosity at the jet break and the X-ray luminosity, and constrain the likely contribution the synchrotron jet has to the observed X-ray luminosity.

In Fig. 3 the jet break frequency $\nu_{\rm b}$ is plotted against peak jet monochromatic luminosity (left panel) and bolometric luminosity (in Eddington units; right panel) for each constraint discussed in Section 2. We define the monochromatic luminosity as the luminosity in erg s$^{-1}$ divided by the frequency (i.e. flux density scaled for distance); $L_{\nu} = L / \nu \propto F_{\nu} D^2$. The fairly large errors in $L_{\nu}$ and $L_{\rm X} / L_{\rm Edd}$ for some sources are a result of the uncertainties in the distance (and BH mass which is required to calculate $L_{\rm Edd}$) but are necessary to include $L_{\nu}$  to compare between sources. 4U 0614+09 is also included in Fig. 3 in order to compare BHXBs to a neutron star source.

\subsection{Jet break distribution}

The range of jet break frequencies spans more than two orders of magnitude, even at similar jet (and bolometric) luminosities. The jet break with the lowest frequency is that of XTE J1118+480 during its 2005 outburst; $\nu_{\rm b} = (4.5 \pm 0.8) \times 10^{12}$ Hz at $L_{\rm X} = 10^{-3} L_{\rm Edd}$. V4641 Sgr on the other hand, has a jet break at a frequency of $\nu_{\rm b} > 4.7 \times 10^{14}$ Hz, at $L_{\rm X} \sim 0.1 ~ L_{\rm Edd}$.

The jet break can also shift in frequency by more than $\sim 1$ order of magnitude for a single BHXB. It was established that this occurs on timescales of hours in GX 339--4 \citep{gandet11}. Here we report a shift in the jet break in XTE J1118+480 from $\nu_{\rm b} = (2.8 \pm 0.6) \times 10^{13}$ Hz at one epoch in its 2000 outburst, to $\nu_{\rm b} = (4.5 \pm 0.8) \times 10^{12}$ Hz during its 2005 outburst at a very similar X-ray luminosity. The distribution of jet breaks peaks in the NIR, with $\sim 7$ out of $\sim 15$ data points in the range $\log ~\nu_{\rm b} / \rm{Hz} = 14.0$ -- 14.5 (these values are approximate because of the uncertainties from the sources with upper or lower limits on $\nu_{\rm b}$). It is also worth noting that more than half of the 13
epochs for GX 339--4 were red (optically thin) with WISE in 2010 \citep{gandet11}, suggesting a
break in the mid-IR, at frequencies probably around or below $\nu_{\rm b} \sim 1.1 \times 10^{13}$ Hz. The one neutron star XB has a jet break in amongst the distribution, but at slightly lower frequencies than the average BHXB. Due to small number statistics we cannot say if neutron star XB jet breaks are at lower frequencies than those of BHXBs. If jet breaks are found in more neutron star XBs then a statistical comparison can be conducted.

\subsection{Relation between jet break and luminosity}

\subsubsection{Theoretical prediction}

Standard theory of compact jets, applicable to both XBs and AGN \citep[e.g.][]{blanko79}, predicts a positive relation between the jet break frequency and luminosity, and a negative relation between jet break frequency and black hole mass \citep{falcbi95,heinsu03,market03,falcet04,miglet06,coriet09,miglet10,rahoet11} if all other parameters are unchanged. Analytically, under some assumptions (see below) the jet break frequency is expected to scale with jet power (i.e., total power contained in the jet, including radiative and kinetic power) as a power law relation; $\nu_{\rm b} \propto Q_{\rm jet}^{2/3}$ and the jet luminosity in the flat, self-absorbed part of the spectrum seen by the observer scales with the jet power as $L_{\rm \nu ,jet} \propto Q_{\rm jet}^{17/12}$ \citep{falcbi95,market03,falcet04}, resulting in the following relation:
\begin{eqnarray}
  \nu_{\rm b} \propto L_{\rm \nu ,jet}^{8/17}.
\end{eqnarray}

An illustration of how the jet break changes with luminosity and with BH mass is shown in fig. 2 of \cite{falcet04}. These relations result from the derived dependency of the jet flux and jet break frequency on the mass accretion rate $\dot{m}$. A constant fraction of the accreted mass is assumed to be channelled into the jets \citep[$Q_{\rm jet} \propto \dot{m}$; see also][]{kordet06,miglet10,rahoet11}:
\begin{equation}
  L_{\rm \nu ,jet} \propto \dot{m}^{17/12},
  \end{equation}

and
\begin{equation}
  \nu_{\rm b} \propto \dot{m}^{2/3}.
\end{equation}

In these models the radiative efficiency of the jet is assumed not to vary with luminosity, and the hard state is considered to be radiatively inefficient, with $L_{\rm X} \propto \dot{m}^2$ (this is appropriate for the direct jet synchrotron model and original advection dominated accretion flow model; see Section 3.23 for alternatives). Thus,
\begin{equation}
  \nu_{\rm b} \propto L_{\rm X}^{1/3}.
\end{equation}

It is important to note at this stage that $\nu_{\rm b}$ is also dependent on the magnetic field strength and the radius of the FAZ, both of which may differ between sources. This is discussed further in Section 3.2.3.

A detailed analytical model is presented by \cite{heinsu03}. The jet break frequency depends on both the mass accretion rate and the lepton energy distribution in the flow, $p$ (see their equation 14 and their following discussion):
\begin{eqnarray}
  \nu_{\rm b} \propto \dot{m}^{\frac{p+6}{2(p+4)}}.
\end{eqnarray}

Similarly, at a constant $\dot{m}$, $\nu_{\rm b}$ depends on both BH mass and $p$ :
\begin{eqnarray}
  \nu_{\rm b} \propto M_{\rm BH}^{-\frac{p+2}{p+4}}.
\end{eqnarray}

From equation 13 of \cite{heinsu03}, if the jet power is a constant fraction of the accretion power then:
\begin{eqnarray}
  L_{\rm \nu ,jet} \propto \dot{m}^{(\frac{17}{12} - \frac{2 \alpha_{\rm thick}}{3})} M_{\rm BH}^{(\frac{17}{12} + \frac{\alpha_{\rm thick}}{3})},
\end{eqnarray}

where $\alpha_{\rm thick}$ is the spectral index of the self-absorbed (optically thick) synchrotron spectrum. Note that here we retain the $F_{\nu} \propto \nu^{\alpha}$ nomenclature, which is not to be confused with that in \cite{heinsu03} in which they adopt $F_{\nu} \propto \nu^{-\alpha}$. From equations (5), (6) and (7) the following relations can be derived \citep[see also][]{coriet09}. At a constant black hole mass but changing mass accretion rate:
\begin{eqnarray}
  \nu_{\rm b} \propto L_{\rm \nu ,jet}^{\frac{6}{17 - 8\alpha_{\rm thick}}\frac{p+6}{p+4}},
\end{eqnarray}

and at a constant mass accretion rate but a changing black hole mass:
\begin{eqnarray}
  \nu_{\rm b} \propto L_{\rm \nu ,jet}^{\frac{-6}{17 + 4\alpha_{\rm thick}}\frac{p+2}{p+4}}.
\end{eqnarray}

The lepton energy distribution can be measured directly from observations, as the spectral index of the optically thin synchrotron emission is directly dependent on $p$; $\alpha_{\rm thin} = (1-p)/2$. In most of the works mentioned above, $p = 2$ is assumed which, when substituted into equation (5) recovers the relation of equation (3). Observationally, $\alpha_{\rm thin}$ has been measured accurately in some X-ray binary jets \citep[e.g.][see also Sections 1 and 2]{hyneet06,miglet10,russet10} and generally resides in the range $\alpha_{\rm thin} \sim -1.0$ -- $-0.5$, which implies $2.0 \leq p \leq 3.0$. Likewise, for a purely flat spectrum, $\alpha_{\rm thick} = 0$ (which is normally assumed) but observationally, radio SEDs of BHXB compact jets have spectral indices $\alpha_{\rm thick} \approx 0.0$ -- +0.5 \citep[e.g.][see also Section 2]{hanhj92,shraet94,fend01}. We arrive at the following ranges of power law indices for the scaling relations, where the ranges encompass all possible values adopting these observational ranges of $\alpha_{\rm thin}$ and $\alpha_{\rm thick}$. At a constant black hole mass but changing mass accretion rate:
\begin{eqnarray}
  \nu_{\rm b} \propto L_{\rm \nu ,jet}^{0.53 \pm 0.08},
\end{eqnarray}

and at a constant mass accretion rate but a changing black hole mass:
\begin{eqnarray}
  \nu_{\rm b} \propto L_{\rm \nu ,jet}^{-0.23 \pm 0.02}.
\end{eqnarray}

The famous radio--X-ray correlation for hard state BHXBs \citep[e.g.][]{corbet00,gallet03,market03} can be recovered from equations (4) and (10):
\begin{eqnarray}
  L_{\rm \nu ,jet} \propto L_{\rm X}^{0.64 \pm 0.10}.
\end{eqnarray}

This is consistent with the empirical global correlations of BHXBs, but there are also radio-faint systems that appear not to follow the same relation \citep[e.g.][]{gallet06,gallet12}. In an independent model, \cite{jamiet10} arrived at a very similar correlation as equation (10). Here, simulations of internal shocks in jets composed of discrete ejections, which take into account adiabatic energy losses in the jet, were able to reproduce the flat, self-absorbed radio to IR jet spectrum. The time-averaged SED produced by the simulations had a jet break frequency which scales with luminosity as $\nu_{\rm b} \propto L_{\rm \nu ,jet}^{\sim 0.6}$.

\subsubsection{Comparing to the observations}

We can compare equations (10) and (4) directly with the observations of jet breaks as shown in the left and right panels of Fig. 3, respectively. One would only expect these relations to explain the data of all sources if other parameters that affect the jet break are the same for each source (see below). This is unlikely, as it is known that some variable parameters of the inflow (e.g. the disc temperature, disc inner radius) differ between sources at the same luminosity, and some fundamental parameters (e.g. the black hole mass) also differ between sources. Nevertheless, the large range of jet breaks at similar luminosities prevents there being one single relation between jet break frequency and luminosity. In addition, since two sources at $L_{\rm X} < 10^{-7} L_{\rm Edd}$ possess jet break frequencies similar to the broad range seen at $L_{\rm X} > 10^{-3} L_{\rm Edd}$ (right panel of Fig. 3), the predicted positive power law relation between luminosity and jet break frequency appears to be weak in the global sample (although we do not have more than a few data points for each source).

We fit a power law\footnote{We swap the axes in order to perform the fit so that the intercept is not a very small or very large number (we are fitting $\nu_{\rm b}(L_{\rm \nu ,jet})$ as opposed to $L_{\rm \nu ,jet}(\nu_{\rm b})$).} to the BHXB data in the left panel of Fig. 3 (neglecting the data which represent only upper/lower limits on $\nu_{\rm b}$) and arrive at a best fit empirical relation:
\begin{eqnarray}
  \nu_{\rm b} \propto L_{\rm \nu ,jet}^{0.05 \pm 0.11}.
\end{eqnarray}

This is a poor fit due to the two orders of magnitude of scatter in $\nu_{\rm b}$ at a similar luminosity. However this best fit to the sample implies the jet break frequency could be independent of luminosity altogether (i.e., a power law index of zero), and the power law index is $> 3\sigma$ away from the expected relation of $\nu_{\rm b} \propto L_{\rm \nu ,jet}^{0.53 \pm 0.08}$. The best fit is shown as a black line in the left panel of Fig. 3 (with errors in the slope shown as dotted lines). In both panels of the figure the theoretical relations are shown in grey. Since we are fitting a compilation of sources with typically one or two data points per source, we cannot rule out the theoretical predicted relation being true for each source, but we can rule out a global relation applicable to all sources. It may be that some sources could obey the theoretical relation, but have different normalizations. This could be due to different values of the magnetic field strength or the radius of the FAZ at the same luminosity between sources. Other parameters are likely to be changing on short timescales, and the global expected relation appears to be lost in the short term changes. This may be why the jet break of XTE J1118+480 was seen to differ by a factor of 10 in frequency on two dates with very similar luminosities (Fig. 3). In addition, \cite{gandet11} showed that $\nu_{\rm b}$ shifted by a factor of 10 in frequency on hour timescales in GX 339--4 while the X-ray luminosity remained largely unchanged on the same timescale.

We note that the best power law fit to the global compilation of sources is sensitive to the two data points at low luminosity. While the jet break of A0620--00 in quiescence has been observed directly, the jet break of GRO J0422+32 is inferred via interpolation, and its error bars are large because we can only infer limits on the spectral index and normalization of the optically thick spectrum. The large errors in $\nu_{\rm b}$ for GRO J0422+32 are not 1$\sigma$ however (they do not represent one standard deviation assuming a Gaussian distribution of possible values of $\nu_{\rm b}$), and the break must reside in the frequency range indicated, for all possible values of $\alpha_{\rm thick}$ and radio luminosity. More jet breaks identified at low luminosities would be beneficial to better constrain the best fit relation.

\subsubsection{The source(s) of the scatter}

Jet models predict the jet break frequency to depend not only on the mass accretion rate, but also on the BH mass, magnetic field strength in the flow ($B$) and the scale height of the launching region (the FAZ), $R_{\rm FAZ}$. In \cite{gandet11} the hour-timescale variability of the jet break frequency was interpreted as dramatic changes in one or both of these two latter parameters. The exact relations are model-dependent, but are derived from standard jet theory. The relations derived are $\nu_{\rm b} \propto B$ and $\nu_{\rm b} \propto R_{\rm FAZ}^{-1}$ \citep[see e.g. equations (1) and (2) of][]{chatet11}. The magnetic field and the acceleration zone are both also expected to vary with mass accretion rate. Here, we explore a number of possibilities that could change the theoretical global relation.

\emph{BHXBs with different BH masses}:
The BHXBs in our sample will have BHs of differing masses. Due to equation (11), we expect some scatter of the jet break frequency between different sources at the same luminosity if they have different BH masses. Here we assess whether this could explain some of the observed scatter in Fig. 3. All BHXBs in our sample have BH masses between 3 and 30 $M_{\odot}$ (a conservative range which encompasses all error bars in all BHs in Table 1). We can calculate the maximum scatter due to these different BH masses thus. Taking all likely electron energy distribution values ($p = 2$ to $p = 3$), equation (6) gives the maximum difference in $\nu_{\rm b}$ between one BHXB with $M_{\rm BH} = 3 M_{\odot}$ and one with $M_{\rm BH} = 30 M_{\odot}$, of a factor of 5.5. So the different BH masses in our sample can theoretically shift the jet break frequency by no more than a factor of 5.5 (0.7 dex) between two BHs at the same luminosity. This is significant, and some of the scatter may be due to the different BH masses, but this cannot explain the much broader range of jet break frequencies observed at the same luminosity.

\emph{Individual ejections and internal shocks}:
In the hard state, there is strong variability in the accretion flow from hours down to less than seconds \citep[typically $\sim 40$ per cent rms variability in the X-ray luminosity;][]{munoet10}, which is well correlated with the jet IR variability \citep{caseet10}. It has been shown that the jet break shifts on hour timescales in GX 339--4 \citep{gandet11}, which is not as rapid as the accretion rate changes that cause the fast variability. Since the size of the emitting region at the jet break is likely to be light seconds across, these hour-long variations cannot be due to individual plasma ejections. Discrete jet `shells' with different velocities in the flow produce additional synchrotron emission when they collide (aka `internal shocks') and this could be on timescales of hours, but are likely to produce emission at lower frequencies than the jet break frequency, as the plasma cools \citep[see][for simulations of individual ejections and the broadband SEDs they produce]{jamiet10}. Sub-second variability has been detected from the optically thin emission \citep[e.g.][]{caseet10}, which can only come from regions close to the BH, not from collisions downstream. Individual ejections and internal shocks are therefore unlikely to be responsible for the hour-long variability of the jet break seen in GX 339--4, nor the longer timescale global scatter in the jet break frequency.

\emph{$\nu_{\rm b}$ dependent on $B$}:

Most jet models explored predict the magnetic field strength to decrease with distance $z$ along the length of the jet, due to the lateral expansion of the jet along its axis, and magnetic flux conservation.   In the seminal model of compact, self-absorbed jets explored by \cite{blanko79}, a conical jet with constant flow velocity has radius $r(z)$ linearly proportional to $z$, and thus $B \propto z^{\sim -1}$.   The self-absorbed synchrotron spectrum from a given segment of this ideal jet will peak at a frequency that is inversely proportional to $z$  \citep[e.g.][]{blanko79}.   The magnetic field strength will therefore scale linearly with the frequency of the jet break, which represents the peak emission from the smallest radiating scale in the jets.    Furthermore, if the internal pressures depend linearly on the accretion rate, then this break frequency would be expected to scale as $\nu_b\propto \dot{m}^{2/3}$ for sources with the same mass, again for the idealized case \citep[see, e.g.,][]{falcbi95,market03,heinsu03}.

Using such ideal assumptions, \cite{chatet11} derived the relation $B \propto \nu_{\rm b} L_{\rm \nu ,jet}^{-1/9}$ \citep[their equation (1); see also][]{gandet11}. This analytical solution was found by solving for the flux of the synchrotron emission at the jet break as a function of the synchrotron absorption coefficient, the optically thin emissivity and the size of the emitting region \citep{rybili79}, assumed to be a homogeneous, cylindrical jet seen sideways.  Using such relations, the magnetic field strength can be estimated to within a rough factor based on direct measurement of the jet break.

In reality, the exact dependence of $r(z)$ will depend on the balance of the internal and external pressures, and interpretation will thus be model dependent.  For instance, a strongly externally confined jet would be expected to have smaller jet opening angles resulting in a less rapid decrease of $B$ with $z$ \citep[e.g.][]{kais06}. If internal or external pressures vary between sources or in time for a particular source, this could also vary the magnetic field strength and configuration. It is thus not too surprising that the idealized relation $B \propto \nu_{\rm b} L_{\rm \nu ,jet}^{-1/9}$ is hidden in the scatter between sources as shown in Fig. 3, which implies that these sources are not strictly self-similar.  In order to unearth meaningful trends, these data sets must be fit by the same model, to determine whether a consistent interpretation linking the breaks, magnetic fields and source characteristics can be found.   We are planning to carry this out in future work. Essentially, we find that different magnetic field strengths in different sources could reproduce the observed scatter in jet break frequencies seen in Fig. 3.

It is worth noting that in a related work, \cite{peerca09} and \cite{casepe09} consider a jet which is accelerated once at its base; here it was found that above a critical magnetic field strength, $\sim 10^5$ G (this varies depending on other parameters), electrons rapidly cool, producing suppressed radio emission but enhanced optical/IR emission. From constraints on the jet breaks in GX 339--4 and XTE J1550--564, magnetic field strengths of $\sim 1$--$5 \times 10^4$ G were reported \citep{gandet11,chatet11}, slightly below, but close to, this critical value. The scatter in Fig. 3 implies that there is likely to be a large range in magnetic field strengths between sources. For some BHXBs with the higher jet break frequencies the critical value \citep{casepe09} may be reached, implying that a flux enhancement could exist in the optical/IR regime. While no enhancement is seen in our sample at these wavelengths (since we measure IR spectral indices which are negative; see Table 2), this scenario cannot be ruled out if the enhancement is at higher or lower frequencies than those sampled in each SED (see below regarding more complex SEDs).

\emph{$\nu_{\rm b}$ dependent on $R_{\rm FAZ}$}:
\cite{miglet10} point out that the jet break frequency scales as $\nu_{\rm b} \propto \dot{m}^{2/3} R_{\rm FAZ}^{-1}$ (a revised version of equation (3)) and that the radius of the FAZ is dependent on the inner radius of the accretion disc \citep[see also][]{falcet04,chatet11}. The inner radius of the disc ($R_{\rm disc}$) can be constrained from X-ray observations, and it is known that at lower luminosities in the hard state the disc becomes more truncated. Using data from a wealth of BHXBs, \cite{cabaet09} derive an empirical relation between the inner disc radius and the X-ray luminosity in the hard state: $R_{\rm disc} \propto L_{\rm X}^{-1/3}$. If the radius of the FAZ is directly anchored to the inner edge of the disc then one expects $R_{\rm FAZ} \propto R_{\rm disc}$ \citep[see also][]{peerma12}, and hence from equation (4), $\nu_{\rm b} \propto L_{\rm X}^{1/3} R_{\rm FAZ}^{-1} \propto L_{\rm X}^{2/3}$. Taking this dependency of $R_{\rm FAZ}$ on $\dot{m}$ into account results in a steeper relation between jet break frequency and luminosity, and so is unlikely to be the case.

The opening angle of the jet plasma entering the FAZ also affects the radius of the FAZ. This opening angle, and how it varies with mass accretion rate, is unknown, but has been seen to be quite wide ($\sim 60^{\circ}$) in one AGN \citep*{junoet99}.

\emph{An alternative relation proposed for GX 339--4}:
\cite{nowaet05} perform broken power law fits to quasi-simultaneous radio and X-ray data of GX 339--4 in its hard state. The jet break frequency is constrained here by the interpolation of the two power laws, and an empirical correlation is found; $\nu_{\rm b} \propto L_{\rm X}^{0.91}$ (see their fig. 3). This relation is steeper than the theoretical relation in equation (4). \cite{nowaet05} adopt the assumption that the optically thin jet synchrotron emission is the origin of the X-ray power law. This may be the case at some X-ray luminosities in the hard state, but may not be the case for all luminosities. Such a steep relationship extrapolated to low luminosities would imply a jet break in the radio domain or even lower frequency in quiescence; much lower frequencies than observed. Correlations between IR and X-ray luminosities favour a different relationship between jet break frequency and luminosity in GX 339--4 \citep{coriet09}. Multiple detections of the jet break in GX 339--4 over an outburst cycle would be useful to test these, and the theoretical relations.

\emph{Relaxing the assumption of a radiatively inefficient hard state}:
Some BHXBs appear too faint in radio compared to the normal radio--X-ray correlation for hard state BHXBs \citep[e.g.][]{solefe11,gallet12}. Data from several outbursts of the BHXB H1743--322 indicated that at moderate luminosities this is a radio-faint BHXB, but that the radio--X-ray correlation steepened at high luminosities in the hard state to a correlation expected for a radiatively efficient accretion flow \citep{coriet11}. We may therefore consider radiatively efficient accretion flows in the hard state here. In this scenario, $L_{\rm X} \propto \dot{m}$, which leads to $\nu_{\rm b} \propto L_{\rm X}^{2/3}$. This is a much steeper correlation than the observations suggest.

\emph{Relaxing the assumption of a constant jet radiative efficiency}:
If the fraction of jet kinetic energy which is radiated away ($\eta$) varies with accretion rate, this will change the correlations and scaling relations. Specifically, the $L_{\rm \nu ,jet} \propto Q_{\rm jet}^{17/12}$ relation breaks down in this scenario. Let us consider a jet radiative efficiency which scales as a power law relation with mass accretion rate; $\eta \propto \dot{m}^{\beta}$. We therefore have $L_{\rm X} \propto \dot{m}^2$ (unchanged) and $L_{\rm \nu ,jet} \propto \eta \dot{m}^{17/12} \propto \dot{m}^{\beta + (17/12)}$. The revised relations between jet break frequency and luminosity would then be $\nu_{\rm b} \propto L_{\rm X}^{1/3}$ (as before) and $\nu_{\rm b} \propto L_{\rm \nu ,jet}^{\frac{8}{12\beta + 17}}$. We can immediately see that since the $\nu_{\rm b}$--$L_{\rm X}$ relationship is unchanged, the distribution of jet breaks in the right panel of Fig. 3 is not affected by a changing jet radiative efficiency.

If we consider only the $\nu_{\rm b}$--$L_{\rm \nu ,jet}$ relation, using the best fit to the data (equation (13)) we find $\frac{8}{12\beta + 17} \approx 0.05$, hence $\beta \approx 11.9$. This implies an extremely strong dependency of radiative efficiency on mass accretion rate globally, which is likely to be unphysical. It also predicts a very steep correlation between jet and X-ray luminosities, of the form $L_{\rm \nu ,jet} \propto L_{\rm X}^{\sim 7}$ which, as we will see below, is inconsistent with the observations. Different jet radiative efficiencies in different sources could introduce scatter in the $\nu_{\rm b}$--$L_{\rm \nu ,jet}$ plot, but not the $\nu_{\rm b}$--$L_{\rm X}$ plot. Since the observed scatter is similar in both plots (Fig. 3), it is unlikely that this is a main source of the scatter.

\emph{More complex SEDs}:
In order to identify the jet break, for many sources it was necessary to interpolate radio and IR/optical power laws. This assumes the jet spectrum can be described approximately by a broken power law (see Section 2). This is the classical model for compact jets \citep{blanko79,heinsu03,falcet04} but more complex models have been developed. Some models predict an extra excess of emission in the SED at a frequency near the jet break \citep{market05,peerca09}, whereas some argue the flat optically thick spectrum is difficult (but not impossible) to reproduce theoretically, if adiabatic energy losses are taken into account \citep{kais06,jamiet10}.

We note that the jet break directly observed in V404 Cyg actually requires two power laws (or a curved spectrum) between radio frequencies and the jet break. Without the IR data, we would have interpolated the radio and optical power laws assuming one break, and arrived at a jet break $\sim 0.5$ dex lower in frequency than the one observed. This demonstrates some scatter in Fig. 3 may be due to more complex SEDs.

We conclude that variations of the magnetic field strength and/or the radius of the FAZ on timescales generally shorter than the outburst timescale likely introduce some of the scatter in the observed relation between jet break frequency and luminosity. The inner disc radius does not change rapidly so if the FAZ is anchored to the inner disc radius, then it is more likely that the magnetic field is the quantity that is varying rapidly. It may also be that the flat optically thick regime may not be well approximated by a single power law in some cases. In this scenario, interpolation of the optically thick and optically thin power laws may introduce some inaccuracies in our estimations of the jet break, which could introduce more scatter in Fig. 3.

\begin{figure}
\centering
\includegraphics[width=6.0cm,angle=270]{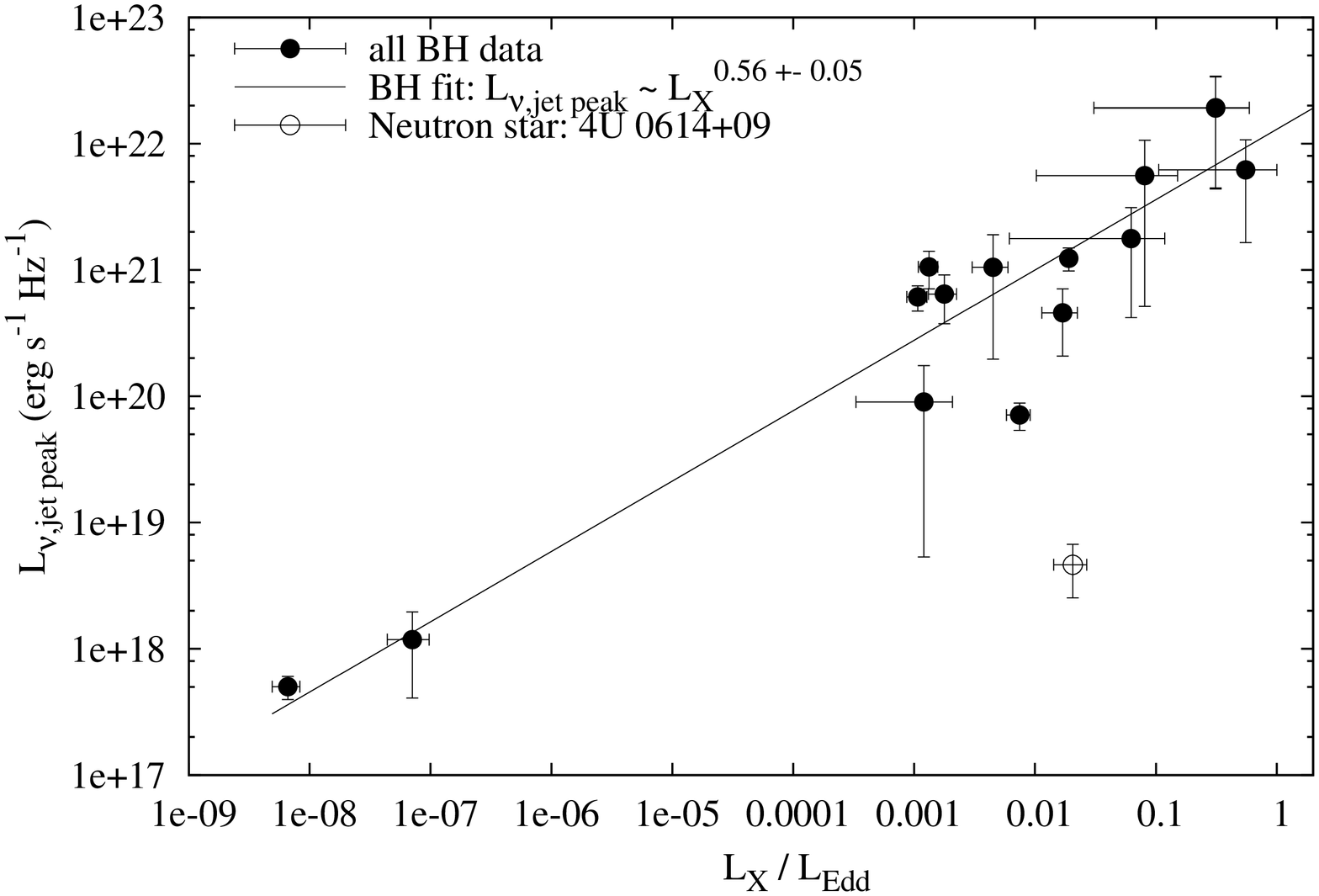}\\
\includegraphics[width=6.0cm,angle=270]{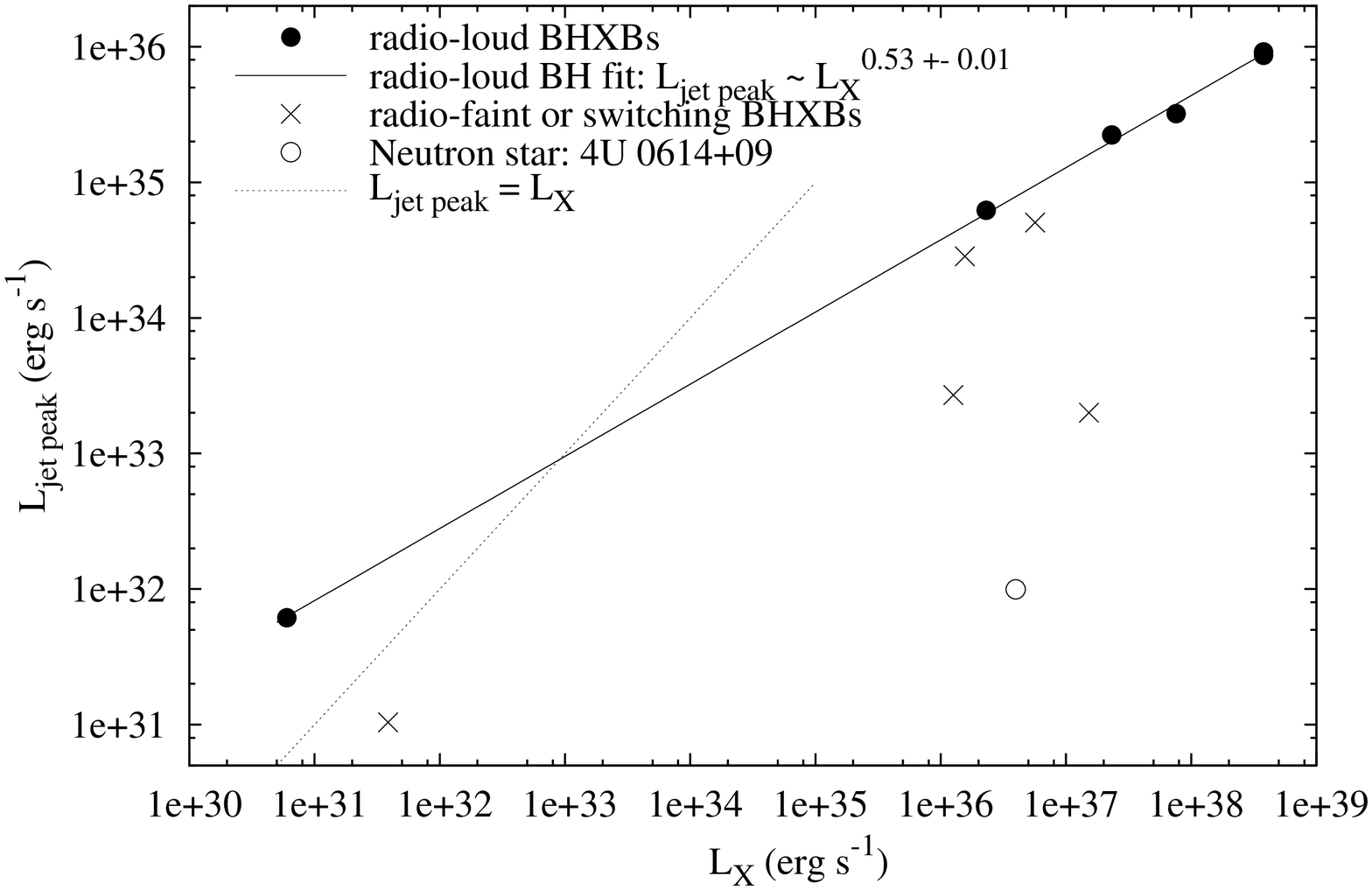}
\caption{\emph{Upper panel:} Monochromatic luminosity of the jet at the jet break frequency versus X-ray luminosity in Eddington units. As well as detections of the jet break, jet break monochromatic luminosities constrained from upper/lower limits on the jet break frequency are also included. \emph{Lower panel:} Luminosity of the jet at the jet break frequency versus X-ray luminosity. Here, only secure detections of the jet break are used so that the most accurate measurements are used only (no upper or lower limits on $\nu_{\rm b}$ are included; this is the reason for fewer data points appearing in the lower panel). The BHXBs that follow the `normal' radio--X-ray correlation (`radio-bright' BHXBs) are shown as separate symbols to BHXBs that appear to be radio-faint compared to this correlation, or appear to lie (or switch) between the two luminosity tracks \citep[e.g.][]{gallet12}.}
\end{figure}

\subsubsection{The radio--X-ray and jet break--X-ray correlations}

The radio--X-ray correlation becomes steeper for a radiatively efficient flow; $L_{\rm \nu ,jet} \propto L_{\rm X}^{1.29 \pm 0.19}$ (where these errors are propagated from those in equation (10)). If a radiatively efficient flow could exist in the hard state, this relation can be tested directly against the observations \citep[see e.g.][for different radio--X-ray slopes found in some BHXBs]{coriet11,gallet12}. In Fig. 4 the observed jet luminosity at the jet break versus the quasi-simultaneous X-ray luminosity is plotted. In the upper panel, monochromatic jet luminosities are used against X-ray luminosities in Eddington units, for all the data considered in this paper. The best fit power law correlation to the BHXB data is $L_{\rm \nu ,jet} \propto L_{\rm X}^{0.56 \pm 0.05}$, clearly inconsistent with a radiatively efficient flow. The flow is consistent with being radiatively inefficient, as the correlation slope lies within the error bars of the expected slope in equation (12). We can also conclude that since $L_{\rm radio} \propto L_{\rm X}^{0.6}$ for the BHXBs that follow the classical radio--X-ray correlation \citep[e.g.][]{gallet06}, the radio luminosity of the jet must scale approximately linearly with the jet luminosity at the jet break; $L_{\rm radio} \propto L_{\rm \nu ,jet}$. Therefore, the spectral index of the self-absorbed synchrotron spectrum, spanning from radio to the jet break, does not vary as a function of luminosity in the hard state.

It is interesting to note that three sources appear too faint in $L_{\rm \nu ,jet}$ in the upper panel of Fig. 4 compared to the correlation. The white circle is the neutron star XB 4U 0614+09. Its jet break luminosity is $> 2$ orders of magnitude less luminous than most BHXBs at the same X-ray luminosity. Radio jets in neutron stars are less luminous than radio jets in BHXBs \citep{miglfe06}. This confirms what was found by \cite{russet07c}; that the same appears to be the case for IR jets; IR jets in neutron stars are fainter than IR jets in BHXBs \citep*[see also][]{miglet11}. The two BHXBs that appear to have fainter jets than the correlation (both are around $L_{\rm X} \sim 0.001$--0.01 $L_{\rm Edd}$) are XTE J1752--223 and Cyg X--1. It is relevant to mention that both sources are (slightly) radio-faint BHXBs \citep{gallet12,rattet12}, so it is not unexpected that their jet breaks are also fainter than most BHXBs.

In the lower panel of Fig. 4, the flux density of the jet at the jet break, scaled to distance (to compare between sources, and approximated by $\nu_{\rm b} L_{\rm \nu ,jet}$; which requires a direct detection of the jet break, not an upper or lower limit on its frequency) is plotted against the bolometric luminosity, both in erg s$^{-1}$. This is a different relation, as $L_{\rm \nu ,jet}$ is not measured at a fixed frequency, but this time is the total luminosity at the jet break, which may be a good indication of the total jet power (at least the total relative jet power between sources). A correlation with a very similar slope as measured in the upper panel of Fig. 4, $L_{\rm jet} \propto L_{\rm X}^{0.53 \pm 0.01}$ is measured from the population of BHXBs that produce the well known radio--X-ray correlation in BHXBs, with very little scatter (black filled circles in the lower panel of Fig. 4). The dotted line in this figure indicates where the jet luminosity equals the X-ray luminosity. It implies that the jet luminosity exceeds the X-ray luminosity below $L_{\rm X} \sim 10^{33}$ erg s$^{-1}$, which is consistent with jet dominated sources existing at low luminosities \citep*[see e.g.][]{fendet03,russet10}.

The BHXBs that appear radio-faint compared to the originally defined radio--X-ray correlation, or sources that appear to move between two luminosity tracks (crosses in Fig. 4), here have fainter jet break luminosities than this correlation \citep[we use the radio--X-ray diagrams of][to define which sources are radio-bright and which are radio-faint]{gallet12,rattet12}. This shows that the two luminosity tracks in the radio--X-ray diagram are actually reproduced in the jet break luminosity--X-ray diagram. The BHXBs which are radio-faint tend to also be IR-faint.

\begin{figure}
\centering
\includegraphics[width=6.1cm,angle=270]{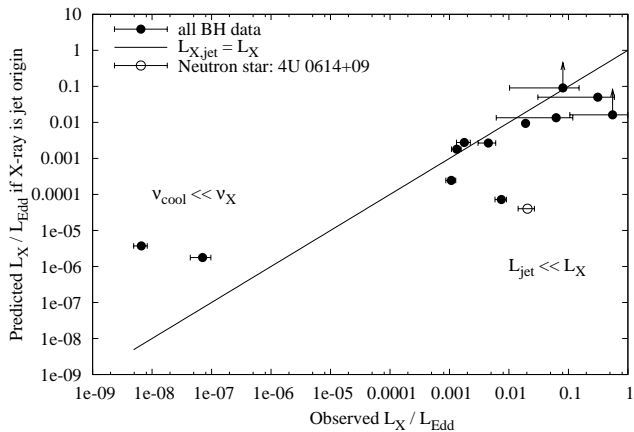}
\caption{Predicted X-ray luminosity of the jet, if the synchrotron power law extends from the observed jet break to X-ray energies (we adopt $F_{\nu} \propto \nu^{-0.8}$) versus the observed X-ray luminosity.}
\end{figure}

\subsubsection{Predicting the X-ray luminosity of the jet}

Since we have measured the jet break frequency and its luminosity, we can directly infer the X-ray luminosity of the jet, under two assumptions. The first is that the optically thin power law extends to X-ray energies (the high energy cutoff, or the cooling break, resides at energies $\geq 10$ keV). This cooling break is hard to detect, but observations have favoured a cooling break in the energy range $\sim 2$--50 keV \citep[see][for a recent discussion]{peerma12}. The second assumption is that of the spectral index of the optically thin power law. For each measurement of the jet break, we compute the X-ray (2--10 keV) luminosity of the extrapolated optically thin jet spectrum assuming its spectral index is $\alpha_{\rm thin} = -0.8$ (a fairly typical spectral index). Adopting different values of $\alpha_{\rm thin}$ would shift each predicted X-ray luminosity systematically by the same amount in log($L_{\rm X}$).

Fig. 5 shows the predicted X-ray luminosity of the optically thin jet power law plotted against the observed X-ray luminosity. For all BHXBs except those in quiescence, the predicted jet luminosity is approximately equal to, or less than the observed X-ray luminosity. Above the quiescent luminosity, all BHXBs have X-ray luminosities that are no brighter than $\sim 10$ times the predicted jet luminosity. The only exceptions are Cyg X--1, which is $\sim 100$ times more luminous, and the neutron star XB 4U 0614+09. It has been shown that in both of these objects, the X-ray emission is very unlikely to originate from synchrotron emission from the jet \citep{miglet10,malzet09}. Fig. 5 confirms this for these two sources.

For most BHXBs, the jet luminosity underpredicts the observed X-ray luminosity by a factor of up to ten, implying that the jet does not dominate the X-ray emission. This is consistent with the standard picture of the Comptonized corona producing the X-ray power law in the hard state \cite[see e.g.][for a review]{gilf09}. We have here showed however that the jet emission could typically contribute more than 10 per cent of the X-ray flux, if the cooling break does not reside at lower energies than the X-ray band.

At $L_{\rm X} \sim 10^{-3} L_{\rm Edd}$, three sources (XTE J1118+480, 4U 1543--47 and XTE J1550--564) have observed X-ray luminosities within a factor of two of the predicted jet luminosities. Empirical evidence has favoured a jet origin to the X-ray emission at this luminosity in XTE J1550--564 \citep{russet10}. XTE J1118+480 was the first BHXB that was proposed to have a jet dominating the X-ray luminosity, and this was at a similar X-ray luminosity \citep{market01}. Fig. 5 suggests that the jet of 4U 1543--47 may also have dominated its X-ray luminosity at $L_{\rm X} \sim 10^{-3}$. The data are broadly consistent with the suggestion that the jet does not dominate the X-ray luminosity above $\sim 2 \times 10^{-3}$ $L_{\rm Edd}$, but may do below that luminosity \citep{russet10}.

The two quiescent BHXBs have observed X-ray luminosities that are several orders of magnitude (three in the case of A0620--00) fainter than the predicted jet luminosity. For this to occur, the cooling break in the jet spectrum must lie at energies much lower than the X-ray regime ($\nu_{\rm cool } \ll \nu_{\rm X}$). It is possible that the optically thin spectral index becomes steeper (e.g. $\alpha_{\rm thin} < -1.0$) at lower luminosities. A correlation was found between $\alpha_{\rm thin}$ and luminosity for XTE J1550--564, in which it was as steep as $\alpha_{\rm thin} = -1.5$ at the lowest luminosities as the jet first appeared after the soft-to-hard state transition \citep{russet10}.

It is also interesting that the measured X-ray spectral index of quiescent BHXBs is generally steeper ($-3 \leq \alpha \leq -2$) than is typical for the hard state \citep[e.g.][]{corbet06}. Indeed, the observed spectral indices in quiescence are consistent with that expected from the jet at energies higher than the cooling break \citep[e.g.][]{peerma12}. Perhaps this implies that the origin of the X-ray emission in quiescent BHXBs is the jet \citep[as predicted by, e.g.][]{yuancu05}, but this is speculative. We can conclude that it is likely that the high energy break in the jet spectrum shifts from X-ray energies at $\sim 10^{-3} L_{\rm Edd}$ to UV energies at $\sim 10^{-8} L_{\rm Edd}$.

\section{Summary and conclusions}

We have collected multiwavelength SEDs of BHXBs in their hard X-ray states. Synchrotron emission from the jets launched in these systems is identified spectrally, and constraints on the frequency and luminosity of the characteristic break in the jet spectrum have been achieved for a total of eight BHXBs. We detect the jet break directly in the SED of V404 Cyg during its 1989 outburst, at $\nu_{\rm b} = (1.8 \pm 0.3) \times 10^{14}$ Hz ($1.7 \pm 0.2 \mu m$). Jet break frequencies span more than two orders of magnitude, and there appears to be no strong global relation between jet break frequency and luminosity from $L_{\rm X} \sim 10^{-8}$--1 $L_{\rm Edd}$. In two sources, GX 339--4 and XTE J1118+480, the jet break frequency varies by more than one order of magnitude while the change in luminosity is negligible \citep[seen on hour-timescales by][in the case of GX 339--4]{gandet11}.

The wide scatter in the relation between jet break frequency and luminosity may be due to the magnetic field strength and/or the radius of the launching region in the jet varying dramatically, causing large shifts in the jet break frequency, and suppressing the visibility of any global relation with luminosity. Different BH masses in different sources are unlikely to cause much of the scatter, but differences in the magnetic field strength, radius of the FAZ, or the jet radiative efficiency could contribute to the scatter in jet break frequencies. The high energy (cooling) break in the jet spectrum appears to shift in frequency, from UV energies at $\sim 10^{-8} L_{\rm Edd}$ (quiescence) to X-ray energies at $\sim 10^{-3} L_{\rm Edd}$.

We find a power law correlation between the jet peak flux (i.e., at the jet break frequency) and the X-ray luminosity for hard state BHXBs; $L_{\rm \nu ,jet} \propto L_{\rm X}^{0.56 \pm 0.05}$; very similar to the well documented radio--X-ray correlation. This implies a radiatively inefficient accretion flow in the hard state; $L_{\rm X} \propto \dot{m}^{\sim 2}$. The BHXBs that are radio-faint in the radio--X-ray correlation (or perform transitions between being radio-faint and radio-bright) are also IR-faint, and the radio to IR spectral index of the jet is independent of luminosity.

\section*{Acknowledgments}

DMR acknowledges support partly from a Netherlands Organisation for Scientific Research (NWO) Veni Fellowship and partly from a Marie Curie Intra European Fellowship within the 7th European Community Framework Programme under contract no. IEF 274805. DMR and TS acknowledge support from the Spanish Ministry of  Science and Innovation (MICINN) under the grant AYA2010-18080. SM gratefully acknowledges support from a Netherlands Organisation for Scientific Research (NWO) Vidi Fellowship, and from the European Community's Seventh Framework Program (FP7/2007-2013) under grant agreement number ITN 215212 `Black Hole Universe'.

\label{lastpage}


\begin{thebibliography}{99}
\bibitem[\protect\citeauthoryear{Belloni}{2010}]{bell10}Belloni T. M., 2010, in `The Jet Paradigm - From Microquasars to Quasars', ed. T. Belloni, Lect. Notes Phys. 794
\bibitem[\protect\citeauthoryear{Blandford \& Konigl}{1979}]{blanko79}Blandford R. D., Konigl A., 1979, ApJ, 232, 34
\bibitem[\protect\citeauthoryear{Blandford \& Payne}{1982}]{blanpa82}Blandford R. D., Payne D. G., 1982, MNRAS, 199, 883
\bibitem[\protect\citeauthoryear{Blandford \& Znajek}{1977}]{blanzn77}Blandford R. D., Znajek R. L., 1977, MNRAS, 179, 433
\bibitem[\protect\citeauthoryear{Brocksopp et al.}{2001}]{brocet01}Brocksopp C., Jonker P. G., Fender R. P., Groot P. J., van der Klis M., Tingay S. J., 2001, MNRAS, 323, 517
\bibitem[\protect\citeauthoryear{Brocksopp, Bandyopadhyay \& Fender}{Brocksopp et al.}{2004}]{brocet04}Brocksopp, C., Bandyopadhyay, R. M., Fender, R. P. 2004, NewA, 9, 249
\bibitem[\protect\citeauthoryear{Brocksopp et al.}{2010}]{brocet10}Brocksopp C., Jonker P. G., Maitra D., Krimm H. A., Pooley G. G., Ramsay G., Zurita C., 2010, MNRAS, 404, 908
\bibitem[\protect\citeauthoryear{Buxton \& Bailyn}{2004}]{buxtba04}Buxton, M. M., Bailyn, C. D. 2004, ApJ, 615, 880
\bibitem[\protect\citeauthoryear{Buxton et al.}{2012}]{buxtet12}Buxton M. M., Bailyn C. D., Capelo H. L., Chatterjee R., Din\c{c}er T., Kalemci E., Tomsick J. A., 2012, AJ, 143, 130
\bibitem[\protect\citeauthoryear{Cabanac et al.}{2009}]{cabaet09}Cabanac C., Fender R. P., Dunn R. J. H., K\"ording E. G., 2009, MNRAS, 396, 1415
\bibitem[\protect\citeauthoryear{Cadolle Bel et al.}{2011}]{cadoet11}Cadolle Bel M., et al., 2011, A\&A, 534, A119
\bibitem[\protect\citeauthoryear{Cantrell et al.}{2010}]{cantet10}Cantrell A. G., et al., 2010, ApJ, 710, 1127
\bibitem[\protect\citeauthoryear{Cardelli, Clayton \& Mathis}{Cardelli et al.}{1989}]{cardet89}Cardelli J. A., Clayton G. C., Mathis J. S., 1989, ApJ, 345, 245
\bibitem[\protect\citeauthoryear{Casares et al.}{1991}]{casaet91}Casares J., Charles P. A., Jones D. H. P., Rutten R. G. M., Callanan P. J., 1991, MNRAS, 250, 712
\bibitem[\protect\citeauthoryear{Casares et al.}{2004}]{casaet04}Casares J., Zurita C., Shahbaz T., Charles P. A., Fender R. P. 2004, ApJ, 613, L133
\bibitem[\protect\citeauthoryear{Casares et al.}{2009}]{casaet09}Casares J., et al., 2009, ApJS, 181, 238
\bibitem[\protect\citeauthoryear{Casella \& Pe'er}{2009}]{casepe09}Casella P., 	Pe'er, A., 2009, ApJ, 703, L63
\bibitem[\protect\citeauthoryear{Casella et al.}{2010}]{caseet10}Casella P., et al., 2010, MNRAS, 404, L21
\bibitem[\protect\citeauthoryear{Castro-Tirado et al.}{1997}]{castet97}Castro-Tirado A. J., Ilovaisky S., Pedersen H., Gonzalez J.-F., Pizarro M., Miranda J., Boehnhardt H., 1997, IAUC, 6775
\bibitem[\protect\citeauthoryear{Chaty et al.}{2002}]{chatet02}Chaty S., Mirabel I. F., Goldoni P., Mereghetti S., Duc P.-A., Mart$\acute{i}$, J., Mignani R. P., 2002, MNRAS, 331, 1065
\bibitem[\protect\citeauthoryear{Chaty et al.}{2003a}]{chatet03a}Chaty S., Haswell C. A., Malzac J., Hynes R. I., Shrader C. R., Cui W., 2003a, MNRAS, 346, 689
\bibitem[\protect\citeauthoryear{Chaty et al.}{2003b}]{chatet03b}Chaty S., Charles P. A., Mart\'i J., Mirabel I. F., Rodr\'iguez L. F., Shahbaz T., 2003b, MNRAS, 343, 169
\bibitem[\protect\citeauthoryear{Chaty, Dubus \& Raichoor}{Chaty et al.}{2011}]{chatet11}Chaty S., Dubus G., Raichoor A., 2011, A\&A, 529, 3
\bibitem[\protect\citeauthoryear{Corbel \& Fender}{2002}]{corbfe02}Corbel S., Fender R. P., 2002, ApJ, 573, L35
\bibitem[\protect\citeauthoryear{Corbel et al.}{2000}]{corbet00}Corbel S., Fender R. P., Tzioumis A.K., Nowak M., McIntyre V., Durouchoux P., Sood R., 2000, A\&A, 359, 251
\bibitem[\protect\citeauthoryear{Corbel et al.}{2001}]{corbet01}Corbel S., et al., 2001, ApJ, 554, 43
\bibitem[\protect\citeauthoryear{Corbel, Tomsick \& Kaaret}{Corbel et al.}{2006}]{corbet06}Corbel S., Tomsick J. A., Kaaret P., 2006, ApJ, 636, 971
\bibitem[\protect\citeauthoryear{Coriat et al.}{2009}]{coriet09}Coriat M., Corbel S., Buxton M. M., Bailyn C. D., Tomsick J. A., K\"ording E., Kalemci E., 2009, MNRAS, 400, 123
\bibitem[\protect\citeauthoryear{Coriat et al.}{2011}]{coriet11}Coriat M., et al., 2011, MNRAS, 414, 677
\bibitem[\protect\citeauthoryear{Curran et al.}{2011}]{curret11}Curran P. A., Maccarone T. J., Casella P., Evans P. A., Landsman W., Krimm H. A., Brocksopp C., Still M., 2011, MNRAS, 410, 541
\bibitem[\protect\citeauthoryear{Din\c{c}er et al.}{2012}]{dincet12}Din\c{c}er T., Kalemci E., Buxton M. M., Bailyn C. D., Tomsick J. A., Corbel S., 2012, ApJ, 753, 55
\bibitem[\protect\citeauthoryear{Eikenberry et al.}{1998}]{eikeet98}Eikenberry S. S., Matthews K., Morgan E. H., Remillard R. A., Nelson R. W., 1998, ApJ, 494, L61
\bibitem[\protect\citeauthoryear{Falcke \& Biermann}{1995}]{falcbi95}Falcke H., Biermann P. L., 1995, A\&A, 293, 665
\bibitem[\protect\citeauthoryear{Falcke, K\"ording \& Markoff}{Falcke et al.}{2004}]{falcet04}Falcke H., K\"ording E., Markoff S., 2004, A\&A, 414, 895
\bibitem[\protect\citeauthoryear{Fender}{2001}]{fend01}Fender R. P., 2001, MNRAS, 322, 31
\bibitem[\protect\citeauthoryear{Fender}{2006}]{fend06}Fender R. P., 2006, in \emph{Compact Stellar X-Ray Sources}, Cambridge University Press, Cambridge, U.K., pp 381 (eds. Lewin W. H. G., van der Klis M.)
\bibitem[\protect\citeauthoryear{Fender et al.}{1997}]{fendet97}Fender R. P., Pooley G. G., Brocksopp C., Newell S. J., 1997, MNRAS, 290, L65
\bibitem[\protect\citeauthoryear{Fender et al.}{2000}]{fendet00}Fender R. P., Pooley G. G., Durouchoux P., Tilanus R. P. J., Brocksopp C., 2000, MNRAS, 312, 853
\bibitem[\protect\citeauthoryear{Fender et al.}{2001}]{fendet01}Fender R. P., Hjellming R. M., Tilanus R. P. J., Pooley G. G., Deane J. R., Ogley R. N., Spencer R. E., 2001, MNRAS, 322, L23
\bibitem[\protect\citeauthoryear{Fender, Gallo \& Jonker}{Fender et al.}{2003}]{fendet03}Fender R. P., Gallo E., Jonker P. G., 2003, MNRAS, 343, L99
\bibitem[\protect\citeauthoryear{Fender et al.}{2009}]{fendet09}Fender R. P., Russell D. M., Knigge C., Soria R., Hynes R. I., Goad M., 2009, MNRAS, 393, 1608
\bibitem[\protect\citeauthoryear{Gallo, Fender \& Pooley}{Gallo et al.}{2003}]{gallet03}Gallo, E., Fender, R. P., Pooley, G. G. 2003, MNRAS, 344, 60
\bibitem[\protect\citeauthoryear{Gallo et al.}{2005}]{gallet05}Gallo E., Fender R. P., Kaiser C., Russell D., Morganti R., Oosterloo T., Heinz S., 2005, Nature, 436, 819
\bibitem[\protect\citeauthoryear{Gallo et al.}{2006}]{gallet06}Gallo E., et al., 2006, MNRAS, 370, 1351
\bibitem[\protect\citeauthoryear{Gallo et al.}{2007}]{gallet07}Gallo E., Migliari S., Markoff S., Tomsick J. A., Bailyn C. D., Berta S., Fender R., Miller-Jones J. C. A., 2007, ApJ, 670, 600
\bibitem[\protect\citeauthoryear{Gallo et al.}{2012}]{gallet12}Gallo E., Miller B. P., Fender R., 2012, MNRAS, 23, 590
\bibitem[\protect\citeauthoryear{Gandhi et al.}{2010}]{gandet10}Gandhi P., et al., 2010, MNRAS, 407, 2166
\bibitem[\protect\citeauthoryear{Gandhi et al.}{2011}]{gandet11}Gandhi P., et al., 2011, ApJ, 740, L13
\bibitem[\protect\citeauthoryear{Gehrz, Johnson \& Harrison}{Gehrz et al.}{1989}]{gehret89}Gehrz R. D., Johnson J., Harrison T., 1989, IAUC, 4816
\bibitem[\protect\citeauthoryear{Gelino \& Harrison}{2003}]{geliha03}Gelino D. M., Harrison T. E., 2003, ApJ, 599, 1254
\bibitem[\protect\citeauthoryear{Gelino et al.}{2006}]{geliet06}Gelino D. M., Balman \c{S}., K{\i}z{\i}lo\u{g}lu \"{U}., Y{\i}lmaz A., Kalemci E., Tomsick J. A., 2006, ApJ, 642, 438
\bibitem[\protect\citeauthoryear{Gelino, Gelino \& Harrison}{Gelino et al.}{2010}]{geliet10}Gelino D. M., Gelino C. R., Harrison T. E., 2010, ApJ, 718, 1
\bibitem[\protect\citeauthoryear{Gilfanov}{2009}]{gilf09}Gilfanov M., 2009, in `The Jet Paradigm - From Microquasars to Quasars', ed. T. Belloni, Lect. Notes Phys. 794 (arXiv:0909.2567)
\bibitem[\protect\citeauthoryear{Goranskii et al.}{1996}]{goraet96}Goranskii V. P., Karitskaya E. A., Kurochkin N. E., Trunkovskii E. M., 1996, AstL, 22, 371
\bibitem[\protect\citeauthoryear{Han \& Hjellming}{1992}]{hanhj92}Han X., Hjellming R. M., 1992, ApJ, 400, 304
\bibitem[\protect\citeauthoryear{Heinz \& Sunyaev}{2003}]{heinsu03}Heinz S., Sunyaev R. A., 2003, MNRAS, 343, L59
\bibitem[\protect\citeauthoryear{Hjellming \& Johnston}{1988}]{hjeljo88}Hjellming R. M., Johnston K. J., 1988, ApJ, 328, 600
\bibitem[\protect\citeauthoryear{Hjellming et al.}{2000}]{hjelet00}Hjellming R. M., et al., 2000, ApJ, 2000, 544, 977
\bibitem[\protect\citeauthoryear{Homan et al.}{2005}]{homaet05}Homan J., Buxton M., Markoff S., Bailyn C. D., Nespoli E., Belloni T., 2005, ApJ, 624, 295
\bibitem[\protect\citeauthoryear{Hynes}{2005}]{hyne05}Hynes R. I., 2005, ApJ, 623, 1026
\bibitem[\protect\citeauthoryear{Hynes et al.}{2002}]{hyneet02}Hynes R. I., Haswell C. A., Chaty S., Shrader C. R., Cui W., 2002, MNRAS, 331, 169
\bibitem[\protect\citeauthoryear{Hynes et al.}{2003}]{hyneet03}Hynes R. I., et al., 2003, MNRAS, 345, 292
\bibitem[\protect\citeauthoryear{Hynes et al.}{2004}]{hyneet04}Hynes R. I., Steeghs D., Casares J., Charles P. A., O'Brien K., 2004, ApJ, 609, 317
\bibitem[\protect\citeauthoryear{Hynes et al.}{2006}]{hyneet06}Hynes R. I., et al., 2006, ApJ, 651, 401
\bibitem[\protect\citeauthoryear{Hynes et al.}{2009}]{hyneet09}Hynes R. I., Bradley C. K., Rupen M., Gallo E., Fender R. P., Casares J., Zurita C., 2009, MNRAS, 399, 2239
\bibitem[\protect\citeauthoryear{Jain et al.}{2001}]{jainet01}Jain R. K., Bailyn C. D., Orosz J. A., McClintock J. E., Remillard R. A., 2001, ApJ, 554, L181
\bibitem[\protect\citeauthoryear{Jamil, Fender \& Kaiser}{Jamil et al.}{2010}]{jamiet10}Jamil O., Fender R. P., Kaiser C. R., 2010, MNRAS, 401, 394
\bibitem[\protect\citeauthoryear{Johnson, Harrison \& Gehrz}{Johnson et al.}{1989}]{johnet89}Johnson J., Harrison T., Gehrz R. D., 1989, IAUC, 4786
\bibitem[\protect\citeauthoryear{Junor, Biretta \& Livio}{Junor et al.}{1999}]{junoet99}Junor W., Biretta J. A., Livio M., 1999, Nat, 401, 891
\bibitem[\protect\citeauthoryear{Kaaret et al.}{2003}]{kaaret03}Kaaret P., Corbel S., Tomsick J. A., Fender R., Miller J. M., Orosz J. A., Tzioumis A. K., Wijnands R., 2003, ApJ, 582, 945
\bibitem[\protect\citeauthoryear{Kaiser}{2006}]{kais06}Kaiser C. R., 2006, MNRAS, 367, 1083
\bibitem[\protect\citeauthoryear{Kalemci et al.}{2005}]{kaleet05}Kalemci E., et al., 2005, ApJ, 622, 508
\bibitem[\protect\citeauthoryear{Khargharia, Froning \& Robinson}{Khargharia et al.}{2010}]{kharet10}Khargharia J., Froning C. S., Robinson E. L., 2010, ApJ, 716, 1105
\bibitem[\protect\citeauthoryear{King, Harrison \& McNamara}{King et al.}{1996}]{kinget96}King N. L., Harrison T. E., McNamara B. J., 1996, AJ, 111, 1675
\bibitem[\protect\citeauthoryear{Kleinmann, Brecher \& Ingham}{Kleinmann et al.}{1976}]{kleiet76}Kleinmann S. G., Brecher K., Ingham W. H., 1976, ApJ, 207, 532
\bibitem[\protect\citeauthoryear{K\"ording, Fender \& Migliari}{K\"ording et al.}{2006}]{kordet06}K\"ording E., Fender R. P., Migliari S., 2006, MNRAS, 369, 1451
\bibitem[\protect\citeauthoryear{Kuulkers et al.}{1999}]{kuulet99}Kuulkers E., Fender R. P., Spencer R. E., Davis R. J., Morison I., 1999, MNRAS, 306, 919
\bibitem[\protect\citeauthoryear{Kuulkers et al.}{2010}]{kuulet10}Kuulkers E., et al., 2010, A\&A, 514, A65
\bibitem[\protect\citeauthoryear{Leibowitz et al.}{1991}]{leibet91}Leibowitz E. M., Ney A., Drissen L., Grandchamps A., Moffat A. F. J., 1991, MNRAS, 250, 385
\bibitem[\protect\citeauthoryear{Livio, Ogilvie \& Pringle}{Livio et al.}{1999}]{liviet99}Livio M., Ogilvie G. I., Pringle J. E., 1999, ApJ, 512, 100
\bibitem[\protect\citeauthoryear{Maitra et al.}{2009a}]{maitet09a}Maitra D., Markoff S., Brocksopp C., Noble M., Nowak M., Wilms J., 2009a, MNRAS, 398, 1638
\bibitem[\protect\citeauthoryear{Maitra, Markoff \& Falcke}{Maitra et al.}{2009b}]{maitet09b}Maitra D., Markoff S., Falcke H., 2009b, A\&A, 508, L13
\bibitem[\protect\citeauthoryear{Maitra et al.}{2011}]{maitet11}Maitra D., Cantrell A., Markoff S., Falcke H., Miller J., Bailyn C., 2011, Proceedings of IAU Symposium 275 ``Jets at all Scales'', 13-17 September 2010, Buenos Aires, Argentina (arXiv:1010.4296)
\bibitem[\protect\citeauthoryear{Malzac, Belmont \& Fabian}{Malzac et al.}{2009}]{malzet09}Malzac J., Belmont R., Fabian A. C., 2009, MNRAS, 400, 1512
\bibitem[\protect\citeauthoryear{Markoff, Falcke \& Fender}{Markoff et al.}{2001}]{market01}Markoff S., Falcke H., Fender R., 2001, A\&A, 372, L25
\bibitem[\protect\citeauthoryear{Markoff et al.}{2003}]{market03}Markoff, S., Nowak, M., Corbel, S., Fender, R., Falcke, H. 2003, A\&A, 397, 645
\bibitem[\protect\citeauthoryear{Markoff, Nowak \& Wilms}{Markoff et al.}{2005}]{market05}Markoff S., Nowak M. A., Wilms J., 2005, ApJ, 635, 1203
\bibitem[\protect\citeauthoryear{McClintock \& Remillard}{2006}]{mcclet06}McClintock J. E., Remillard R. A., 2006, in Compact Stellar X-Ray Sources, eds. Lewin W. H. G., van der Klis M., Cambridge University Press, p. 157
\bibitem[\protect\citeauthoryear{McKinney, Tchekhovskoy \& Blandford}{McKinney et al.}{2012}]{mckiet12}McKinney J. C., Tchekhovskoy A., Blandford R. D., 2012, MNRAS, 423, 3083
\bibitem[\protect\citeauthoryear{Meier}{2001}]{meie01}Meier D. L., 2001, ApJ, 548, L9
\bibitem[\protect\citeauthoryear{Migliari \& Fender}{2006}]{miglfe06}Migliari S., Fender R. P., 2006, MNRAS, 366, 79
\bibitem[\protect\citeauthoryear{Migliari et al.}{2006}]{miglet06}Migliari S., Tomsick J. A., Maccarone T. J., Gallo E., Fender R. P., Nelemans G., Russell D. M., 2006, ApJ, 643, L41
\bibitem[\protect\citeauthoryear{Migliari et al.}{2007}]{miglet07}Migliari S., et al., 2007, ApJ, 670, 610
\bibitem[\protect\citeauthoryear{Migliari et al.}{2010}]{miglet10}Migliari S., et al., 2010, ApJ, 710, 117
\bibitem[\protect\citeauthoryear{Migliari, Miller-Jones \& Russell}{Migliari et al.}{2011}]{miglet11}Migliari S., Miller-Jones J. C. A., Russell D. M., 2011, MNRAS, 415, 2407
\bibitem[\protect\citeauthoryear{Miller-Jones et al.}{2009}]{millet09}Miller-Jones J. C. A., Jonker P. G., Dhawan V., Brisken W., Rupen M. P., Nelemans G., Gallo E., 2009, ApJ, 706, L230
\bibitem[\protect\citeauthoryear{Miller-Jones et al.}{2011}]{millet11}Miller-Jones J. C. A., Jonker P. G., Maccarone T. J., Nelemans G., Calvelo D. E., 2011, ApJ, 739, L18
\bibitem[\protect\citeauthoryear{Mirabel et al.}{1998}]{miraet98}Mirabel I. F., Dhawan V., Chaty S., Rodriguez L. F., Marti J., Robinson C. R., Swank J., Geballe T., 1998, A\&A, 330, L9
\bibitem[\protect\citeauthoryear{Muno \& Mauerhan}{2006}]{munoma06}Muno M. P., Mauerhan J.,2006, ApJ, 648, L135
\bibitem[\protect\citeauthoryear{Mu\~noz-Darias et al.}{2010}]{munoet10}Mu\~noz-Darias T., Motta S., Pawar D., Belloni T. M., Campana S., Bhattacharya D., 2010, MNRAS, 404, L94
\bibitem[\protect\citeauthoryear{Neilsen \& Lee}{2009}]{neille09}Neilsen J., Lee J. C., 2009, Nat, 458, 481
\bibitem[\protect\citeauthoryear{Nowak et al.}{2005}]{nowaet05}Nowak M. A., Wilms J., Heinz S., Pooley G., Pottschmidt K., Corbel S, 2005, ApJ, 626, 1006
\bibitem[\protect\citeauthoryear{Orosz et al.}{1998}]{oroset98}Orosz J. A., Jain R. K., Bailyn C. D., McClintock J. E., Remillard R. A., 1998, ApJ, 499, 375
\bibitem[\protect\citeauthoryear{Orosz et al.}{2001}]{oroset01}Orosz J. A., et al., 2001, ApJ, 555, 489
\bibitem[\protect\citeauthoryear{Orosz et al.}{2002}]{oroset02}Orosz J. A., Polisensky E. J., Bailyn C. D., Tourtellotte S. W., McClintock J. E., Remillard R. A., 2002, BAAS, 34, 1124
\bibitem[\protect\citeauthoryear{Orosz et al.}{2011a}]{oroset11a}Orosz J. A., Steiner J. F., McClintock J. E., Torres M. A. P., Remillard R. A., Bailyn C. D., Miller J. M., 2011a, ApJ, 730, 75
\bibitem[\protect\citeauthoryear{Orosz et al.}{2011b}]{oroset11b}Orosz J. A., McClintock J. E., Aufdenberg J. P., Remillard R. A., Reid M. J., Narayan R., Gou L., 2011, ApJ, 742, 84
\bibitem[\protect\citeauthoryear{Pavlenko et al.}{2001}]{pavlet01}Pavlenko E. P., Dmitrienko E. S., Shakhovskoi N. M., Shugarov S. Y., Katysheva N. A., Volkov I. M., 2001, ApSSS, 276, 63
\bibitem[\protect\citeauthoryear{Pe'er \& Casella}{2009}]{peerca09}Pe'er A., Casella P., 2009, ApJ, 699, 1919
\bibitem[\protect\citeauthoryear{Pe'er \& Markoff}{2012}]{peerma12}Pe'er A., Markoff S., 2012, ApJ, 753, 177
\bibitem[\protect\citeauthoryear{Polko, Meier \& Markoff}{Polko et al.}{2010}]{polket10}Polko P., Meier D. L., Markoff S., 2010, ApJ, 723, 1343
\bibitem[\protect\citeauthoryear{Ponti et al.}{2012}]{pontet12}Ponti G., Fender R. P., Begelman M. C., Dunn R. J. H., Neilsen J., Coriat M., 2012, MNRAS, 422, L11
\bibitem[\protect\citeauthoryear{Ratti et al.}{2012}]{rattet12}Ratti E. M., et al. 2012, MNRAS, 423, 2656
\bibitem[\protect\citeauthoryear{Rahoui et al.}{2011}]{rahoet11}Rahoui F., Lee J. C., Heinz S., Hines D. C., Pottschmidt K., Wilms J., Grinberg V., 2011, ApJ, 736, 63
\bibitem[\protect\citeauthoryear{Rahoui et al.}{2012}]{rahoet12}Rahoui F., et al., 2012, MNRAS, 422, 2202
\bibitem[\protect\citeauthoryear{Reid et al.}{2011}]{reidet11}Reid M. J., McClintock J. E., Narayan R., Gou L., Remillard R. A., Orosz J. A., 2011, ApJ, 742, 83
\bibitem[\protect\citeauthoryear{Robertson, Warren \& Bywater}{Robertson et al.}{1976}]{robeet76}Robertson B. S. C., Warren P. R., Bywater R. A., 1976, IBVS, 1173, 1
\bibitem[\protect\citeauthoryear{Rupen, Mioduszewski \& Dhawan}{Rupen et al.}{2003}]{rupeet03}Rupen M. P., Mioduszewski A. J., Dhawan V., 2003, Astronomer's Telegram, 172
\bibitem[\protect\citeauthoryear{Russell et al.}{2006}]{russet06}Russell D. M., Fender R. P., Hynes R. I., Brocksopp C., Homan J., Jonker P. G., Buxton M. M., 2006, MNRAS, 371, 1334
\bibitem[\protect\citeauthoryear{Russell et al.}{2007a}]{russet07a}Russell D. M., Fender R. P., Gallo E., Kaiser C. R., 2007a, MNRAS, 376, 1341
\bibitem[\protect\citeauthoryear{Russell et al.}{2007b}]{russet07b}Russell D. M., Maccarone T. J., K\"ording E. G., Homan J., 2007b, MNRAS, 379, 1401
\bibitem[\protect\citeauthoryear{Russell, Fender \& Jonker}{Russell et al.}{2007c}]{russet07c}Russell D. M., Fender R. P., Jonker P. G., 2007c, MNRAS, 379, 1108
\bibitem[\protect\citeauthoryear{Russell \& Fender}{2008}]{russfe08}Russell D. M., Fender R. P., 2008, MNRAS, 387, 713
\bibitem[\protect\citeauthoryear{Russell et al.}{2010}]{russet10}Russell D. M., Maitra D., Dunn R. J. H., Markoff S., 2010, MNRAS, 405, 1759
\bibitem[\protect\citeauthoryear{Russell et al.}{2011a}]{russet11a}Russell D. M., Miller-Jones J. C. A., Maccarone T. J., Yang Y. J., Fender R. P., Lewis F., 2011a, ApJ, 739, L19
\bibitem[\protect\citeauthoryear{Russell et al.}{2011b}]{russet11b}Russell D. M., Casella P., Fender R., Soleri P., Pretorius M. L., Lewis F., van der Klis M., 2011b, in Proceedings of `High Time Resolution Astrophysics IV -- The Era of Extremely Large Telescopes -- HTRA-IV', 5--7 May 2010, Agios Nikolaos, Crete, Greece, Proceedings of Science (arXiv:1104.0837)
\bibitem[\protect\citeauthoryear{Russell et al.}{2011c}]{russet11c}Russell D. M., et al., 2011c, Astronomer's Telegram, 3689
\bibitem[\protect\citeauthoryear{Russell et al.}{2012}]{russet12}Russell D. M., et al., 2012, MNRAS, 419, 1740
\bibitem[\protect\citeauthoryear{Rybicki \& Lightman}{1979}]{rybili79}Rybicki G. B., Lightman A. P., 1979, Radiative Processes in Astrophysics. Wiley, New York
\bibitem[\protect\citeauthoryear{Shahbaz et al.}{2008}]{shahet08}Shahbaz T., Fender R. P., Watson C. A., O'Brien K., 2008, ApJ, 672, 510
\bibitem[\protect\citeauthoryear{Shidatsu et al.}{2011}]{shidet11}Shidatsu M., et al., 2011, PASJ, 63, 785
\bibitem[\protect\citeauthoryear{Shaposhnikov et al.}{2010}]{shapet10}Shaposhnikov N., Markwardt C., Swank J., Krimm H., 2010, ApJ, 723, 1817
\bibitem[\protect\citeauthoryear{Shrader et al.}{1994}]{shraet94}Shrader C. R., Wagner R. M., Hjellming R. M., Han X. H., Starrfield S. G., 1994, ApJ, 434, 698
\bibitem[\protect\citeauthoryear{Soleri \& Fender}{2011}]{solefe11}Soleri P., Fender R, 2011, MNRAS, 413, 2269
\bibitem[\protect\citeauthoryear{Soleri et al.}{2010}]{soleet10}Soleri P., et al., 2010, MNRAS, 406, 1471
\bibitem[\protect\citeauthoryear{Soria, Bessell \& Wood}{Soria et al.}{1997}]{soriet97}Soria R., Bessell M. S., Wood P., 1997, IAUC, 6781
\bibitem[\protect\citeauthoryear{Taranova \& Shenavrin}{2001}]{tarash01}Taranova O. G., Shenavrin V. I., 2001, AstL, 27, 25
\bibitem[\protect\citeauthoryear{Tomsick, Corbel \& Kaaret}{Tomsick et al.}{2001}]{tomset01}Tomsick J. A., Corbel S., Kaaret P., 2001, ApJ, 563, 229
\bibitem[\protect\citeauthoryear{Tomsick et al.}{2003}]{tomset03}Tomsick J. A., Corbel S., Fender R., Miller J. M., Orosz J. A., Tzioumis T., Wijnands R., Kaaret P., 2003, ApJ, 582, 933
\bibitem[\protect\citeauthoryear{Tudose et al.}{2006}]{tudoet06}Tudose V., Fender R. P., Kaiser C. R., Tzioumis A. K., van der Klis M., Spencer R., 2006, MNRAS, 372, 417
\bibitem[\protect\citeauthoryear{Uemura et al.}{2004a}]{uemuet04a}Uemura M., et al., 2004a, PASJ, 56, 61
\bibitem[\protect\citeauthoryear{Uemura et al.}{2004b}]{uemuet04b}Uemura M., et al., 2004b, PASJ, 56, 823
\bibitem[\protect\citeauthoryear{van Paradijs et al.}{1994}]{vanpet94}van Paradijs J., Telesco C. M., Kouveliotou C., Fishman G. J., 1994, ApJ, 429, L19
\bibitem[\protect\citeauthoryear{van Straaten et al.}{2000}]{vanset00}van Straaten S., Ford E. C., van der Klis M., M$\acute{e}$ndez M., Kaaret P. 2000, ApJ, 540, 1049
\bibitem[\protect\citeauthoryear{Veledina, Poutanen \& Vurm}{Veledina et al.}{2011}]{veleet11}Veledina A., Poutanen J., Vurm I., 2011, ApJ, 737, L17
\bibitem[\protect\citeauthoryear{Weingartner \& Draine}{2001}]{weindr01}Weingartner J. C., Draine B. T., 2001, ApJ, 548, 296
\bibitem[\protect\citeauthoryear{Yuan \& Cui}{2005}]{yuancu05}Yuan F., Cui W., 2005, ApJ, 629, 408
\bibitem[\protect\citeauthoryear{Zurita et al.}{2006}]{zuriet06}Zurita C., 2006, ApJ, 644, 432

\end{thebibliography}
\end{document}